\documentclass[sigconf]{acmart}

\renewcommand\footnotetextcopyrightpermission[1]{}

\AtBeginDocument{%
  \providecommand\BibTeX{{%
    \normalfont B\kern-0.5em{\scshape i\kern-0.25em b}\kern-0.8em\TeX}}}

\usepackage{balance}
\usepackage{algorithm}
\usepackage{algorithmic}
\usepackage{multirow}
\usepackage{xspace}
\usepackage{enumitem}
\usepackage{bbding}
\usepackage{stfloats}
\usepackage{makecell}
\usepackage{minted}
\usepackage{tcolorbox}
\usepackage{tabularx}
\usepackage{enumitem}
\usepackage{graphicx}
\usepackage{adjustbox}
\usepackage{multirow}
\usepackage{subcaption}
\usepackage{xcolor}        
\usepackage{colortbl}
\usepackage{amsmath}    
\usepackage{pgfplots}
\pgfplotsset{compat=1.18}

\usepackage{amssymb}
\usepackage{pgfplots}

\def\tool{\texttt{CelerLog}\xspace}

\begin{document}

\title{CelerLog: Fast Log Parsing via Dynamic Routing}

\author{Shiwen Shan$^{\P}$, Yintong Huo$^{\ddag}$, Minxing Wang$^{\ddag}$, Zhiying Wu$^{\P}$, Yuxin Su$^{\P}$, Zibin Zheng$^{\P}$}
\affiliation{%
  \institution{$^{\P}$Sun Yat-sen University, $^{\ddag}$Singapore Management University}
  \country{}
}
\email{{shanshw, wuzhy95}@mail2.sysu.edu.cn}
\email{{ythuo, mxwang}@smu.edu.sg, {suyx35,  zhzibin}@mail.sysu.edu.cn}



\begin{abstract}
Log parsing is a fundamental step for automated log analysis, which transforms raw log messages into structured formats. Existing syntax-based parsers struggle with complex logs because they lack semantic reasoning ability. Emerging LLM-powered semantic parsers achieve high accuracy but suffer from prohibitive latency and token costs because they apply semantic inference across all logs.
Our key observation is that not all logs necessitate complex semantic understanding: a vast majority of logs exhibit repetitive patterns that can be extracted via straightforward statistical analysis. Driven by this insight, we propose \tool, a fast and effective log parser. \tool introduces a dynamic routing mechanism to classify logs into dense and sparse groups. Logs with strong statistical patterns (dense groups) are processed by an efficient statistical processor, whereas the sparse groups lacking such patterns are routed to an LLM for semantic inference.  
This hybrid strategy avoids unnecessary LLM invocations.
Extensive experiments on 14 public datasets show that \tool achieves leading performance over state-of-the-art baselines and is $7.9\times$ to $18.6\times$ faster than LLM methods and up to $1.5\times$ faster than Drain. Additionally, it reduces costs by decreasing token consumption
by 80.2\%--94.1\% and LLM invocations by 86.4\%--90.9\%.
\end{abstract}

\maketitle

\section{Introduction}
\begin{figure}[tbp]
    \centering
    \includegraphics[width=0.9\linewidth]{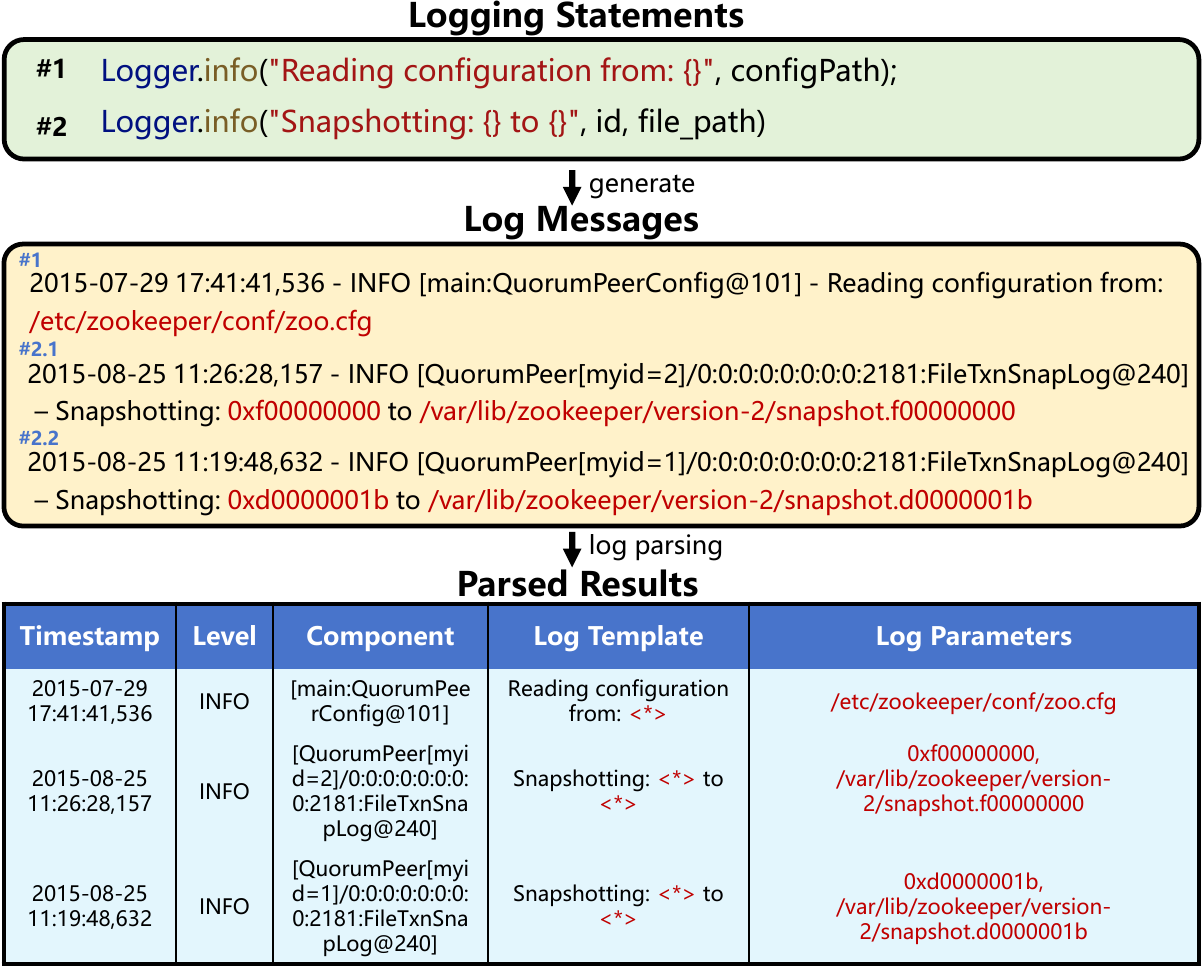}
    \caption{The general paradigm of log parsing.}
    \label{fig:p-log-parsing}
\end{figure}

Log messages are generated at runtime by developer-embedded logging statements to record system behaviors~\cite{schipper2019tracing,jiang2024large,chen2018automated}, serving as a crucial data source for automated tasks such as error diagnosis~\cite{shan2024face,zhou2019latent,yuan2010sherlog,wang2018misconfdoctor} and anomaly detection~\cite{le2021log,du2017deeplog,zhang2019robust,fu2009LKE,shin2021theoretical}. 
As the prerequisite step for these analyses, log parsing transforms semi-structured raw logs into structured formats~\cite{yu2024unlocking,li2024logshrink,vaarandi2015logcluster}. Typically, a log message comprises two main parts: the \textit{log templates} (constant string literals describing event types) and the \textit{log parameters} (dynamic runtime values). Figure~\ref{fig:p-log-parsing} illustrates a workflow: logging statements (e.g., \texttt{Logger.info}) produce raw messages containing headers~(e.g., timestamp) and bodies during execution. The parsing process then identifies variable parameters (e.g., \texttt{/etc/zookeeper/conf/zoo.cfg}) and extracts the static template (e.g., \texttt{Reading configuration from: <*>}) for downstream tasks.

Considering that the source code may not be available, existing parsing tools rely on code-free parsing~\cite{he2016evaluation,zhu2019tools}. Over the decades, log parsing tools can be divided into two categories: syntax-based and semantic-based.
(1) \textit{Syntax-based parsers} employ heuristics or statistical features to identify constant parts from variable ones. 
While generally efficient, these parsers often fall short in parsing effectiveness due to their lack of semantic awareness, struggling to handle complex log patterns accurately.
(2) \textit{Semantic-based parsers}, on the other hand, exploit semantic-aware models to distinguish dynamic parameters from static text. 
Recently, LLM-based parsers have emerged as a dominant solution in this category. By leveraging the powerful natural language understanding capabilities of LLMs, these methods achieve leading parsing accuracy, identifying parameters that traditional syntax rules might miss.

\textit{However, the heavy reliance of these semantic parsers on LLMs introduces prohibitive costs}, such as processing latency and excessive token consumption, limiting their real-world practicality.
Previous works attempted to mitigate this overhead through various strategies. 
For instance, LILAC~\cite{jiang2024lilac} utilizes caching mechanisms to avoid redundant queries, while LUNAR~\cite{huang2025no} and LogBatcher~\cite{xiao2024free} employ clustering to sample representative logs for query. Nevertheless, these methods face a fundamental bottleneck: they still depend on LLMs to process \textit{every} distinct log group, failing to exploit the rich statistical patterns in large-scale log data.

Our investigation reveals a critical insight: \textbf{not all logs require semantic understanding for log parsing}. 
A majority of logs exhibit repetitive patterns with parameter variations, which can be easily extracted using statistical analysis. 
For example, in Figure~\ref{fig:p-log-parsing}, log messages \#2.1 and \#2.2 share the same template with different parameters. When such patterns occur with high frequency, statistical evidence alone is sufficient for automated parsing.
Conversely, only logs with sparse patterns require further semantic reasoning of LLMs.
For instance, log message \#1 appears isolated without repetitive statistical clues. Therefore, it relies on LLMs to understand its semantics and identify its parameters.

Driven by this insight, we propose a fast and effective log parser, namely \tool, with three components: a dynamic \textit{router}, a simple yet fast \textit{statistical processor}, and a semantic-aware \textit{LLM-based processor}. 
The core design is a dynamic router that directs incoming logs to different processors. 
Specifically, it sends \textit{dense log groups} (rich in statistical signals) to the \textit{statistical processor} for instant syntax-based parsing, and routes \textit{sparse log groups} (lacking statistical patterns) to the \textit{LLM-based processor} for semantic analysis.
To ensure efficient routing, we adopt a two-stage grouping phase and an anchor-based merging phase. We first identify skeleton groups and merge similar ones based on shared anchors. Groups that successfully merge with others are classified as \textit{dense log groups}. The \textit{statistical processor} then handles them by identifying columns with multiple distinct values as parameters. The remaining unmerged groups are deemed \textit{sparse log groups}. Since these logs lack statistical variations, the \textit{LLM-based processor} uses its semantic capability to identify variable types within the text. To further reduce costs, we process these sparse groups in batches and enable parallelism to boost speed.

Extensive experiments on 14 public datasets demonstrate the superiority of \tool. 
In terms of effectiveness, \tool outperforms state-of-the-art baselines by an average of 9.6\% in grouping accuracy and 46.8\% in parsing accuracy.
Regarding efficiency, it achieves a $7.9$--$18.6\times$ speedup compared to LLM-based methods, and surpasses the fastest parser, Drain, by up to $1.5 \times$ in the parallel version. Furthermore, \tool significantly reduces costs by decreasing token consumption
by 80.2\%--94.1\% and LLM invocations by 86.4\%--90.9\%.
Further ablation study, robustness analysis and sensitivity analysis confirm \tool's robustness.


To conclude, our contributions are listed as follows:

$\bullet$ We identify that the bottleneck of existing LLM-based parsers is the unified semantic inference procedure to all logs, whereas a majority can be handled by statistical analysis. 

$\bullet$ We introduce the concept of dense and sparse log groups, proposing a hybrid strategy that leverages statistical signals to minimize unnecessary LLM invocations.

$\bullet$ We propose \tool, an innovative log parser featuring a dynamic routing mechanism that directs logs to different processors, thereby maximizing efficiency and effectiveness.

$\bullet$ Comprehensive experiments show that \tool significantly outperforms state-of-the-art baselines in parsing effectiveness, token consumption, and processing time by orders of magnitude.

\begin{figure}[tbp]
    \centering
    \includegraphics[width=\linewidth]{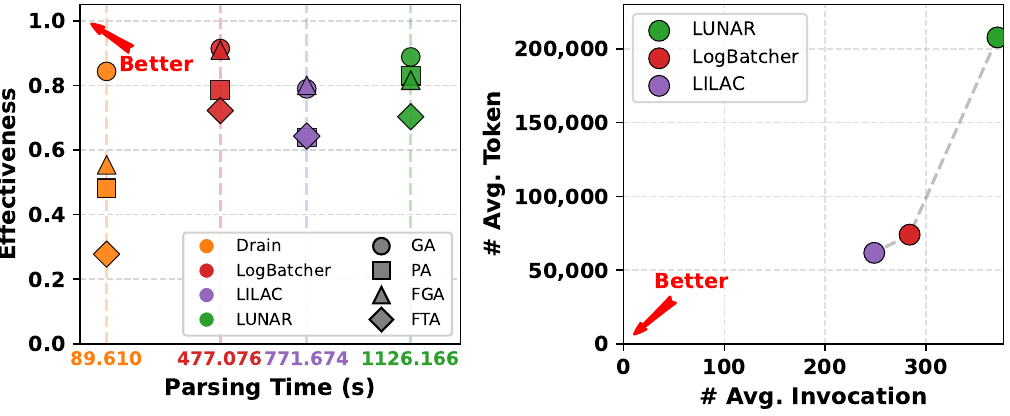}
    \caption{The parsing effectiveness-cost of log parsers. We use GPT-5.2 as the backbone LLM for the LLM-based parsers.}
    \label{fig:comparsion-examples}
\end{figure}
\section{Preliminaries}
In this section, we discuss the trade-off between effectiveness and cost in log parsers, and introduce two distinct log categories: dense and sparse log groups.

\subsection{The Effectiveness-Cost of Log Parsers} 
To comprehensively understand current log parsing tools, we evaluate them through two key dimensions: \textit{parsing effectiveness} and \textit{parsing cost}. Effectiveness refers to the accuracy of identifying log templates and parameters, while cost encompasses both parsing time and the financial expense of LLM invocations.

On one hand, syntax-based parsers prioritize efficiency. For instance, as shown in Figure~\ref{fig:comparsion-examples} (left), Drain~\cite{he2017drain} completes parsing in $89$ seconds on average. Its minimal latency has led to its wide adoption by industry leaders like IBM~\cite{ibm-drain}. However, Drain significantly lags behind LLM-based methods in effectiveness, as it struggles to generalize to diverse or evolving log patterns without semantic understanding.
On the other hand, while LLM-based parsers achieve superior accuracy, they incur prohibitive costs. As illustrated in Figure~\ref{fig:comparsion-examples}, methods like LogBatcher and LUNAR require $477$ to $1,126$ seconds, which is $5\times$ to $13\times$ slower than Drain. 
Beyond latency, the financial burden becomes a concern for deployment. 
In our experiments using GPT-5.2~\cite{gpt}, these methods consume over $50,000$ tokens per dataset on average. 
When scaled to industrial volumes where systems generate petabytes of logs daily~\cite{liu2019logzip,li2024logshrink}, the costs increase linearly and thereby become financially unsustainable.
The root cause of this inefficiency lies in their design principles: existing LLM-based parsers depend heavily on LLM reasoning for every distinct log pattern. Despite optimization strategies like caching or sampling, the fundamental dependency on LLMs for template extraction remains unchanged.

Consequently, although semantic parsers are accurate, this severe effectiveness-cost trade-off hampers their applicability. This observation prompts a critical question: \textit{Is there an approach that bridges semantic and syntax parsers to exploit the strengths of both?}



\begin{figure}[t]
    \centering
    \includegraphics[width=0.5\linewidth]{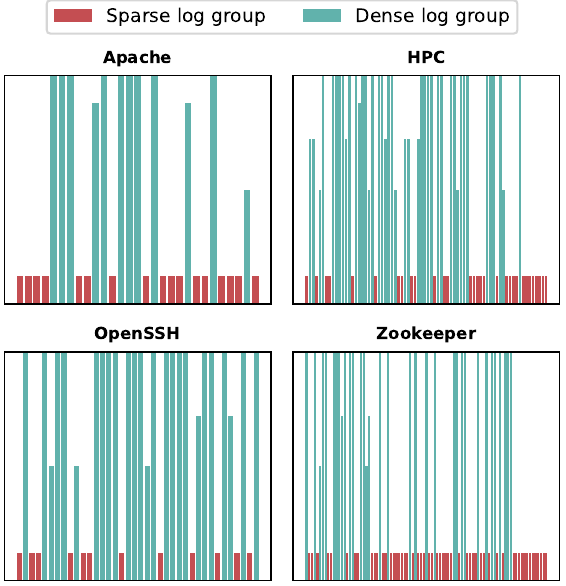}
    \caption{Distribution of dense and sparse log groups across four public datasets. The x-axis denotes unique templates, while the y-axis shows their corresponding message counts. Sparse groups are characterized by a single log message occurrence ($y=1$).}
    \label{fig:two-types-of-log-groups}
\end{figure}
\begin{figure*}[t]
    \centering
    \includegraphics[width=\linewidth]{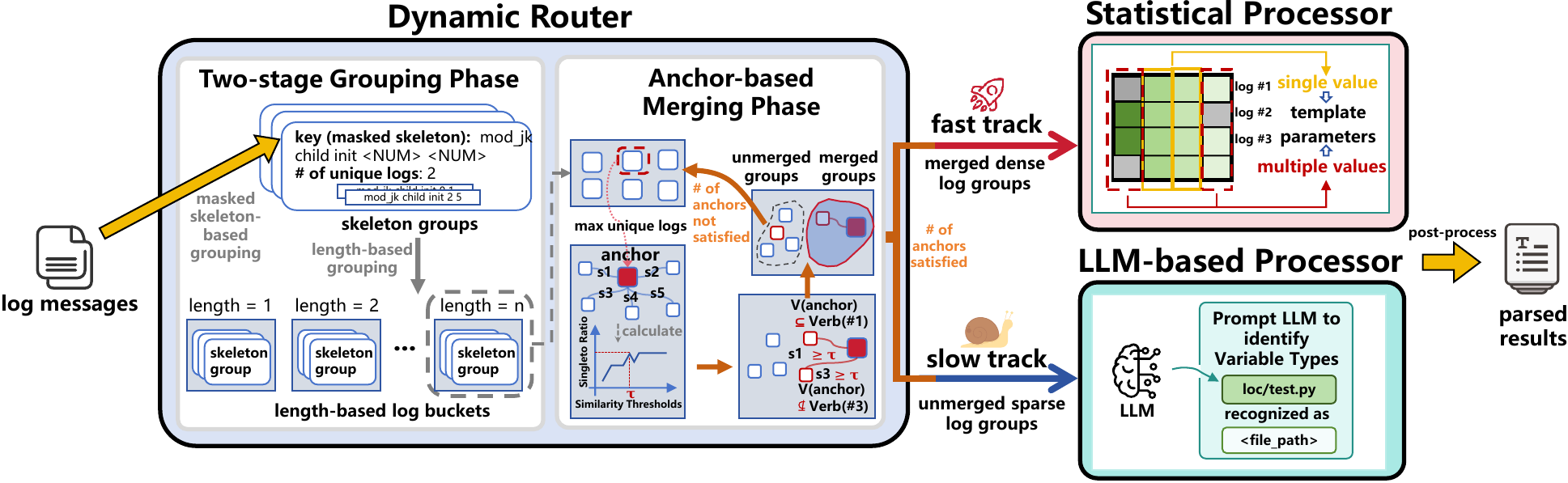}
    \caption{The overview of \tool.}
    \label{fig:overview}
\end{figure*}
\subsection{Dense \& Sparse Log Groups}
To address the above question, we investigate the statistical distribution of log data. 
Our goal is to determine whether there exists a boundary that distinguishes logs suitable for syntax-based parsing from those requiring semantic inference.

\textbf{Inspiration \& Observation.} 
We hypothesize that if a log template generates multiple unique messages with varying parameters, these variations provide sufficient statistical signals for pattern extraction without LLMs. 
Otherwise, the log template lacks such signals and necessitates semantic understanding.
Our empirical analysis on 4 public datasets~(Apache, HPC, OpenSSH and Zookeeper)~\cite{jiang2024large} validates this hypothesis. 
As shown in Figure~\ref{fig:two-types-of-log-groups}, we observe two distinct patterns in log distributions. 
The first type, represented by the tall teal bars, consists of numerous unique log messages derived from the same template. 
The variations in these messages (e.g., changing IP addresses) naturally highlight the parameter positions. 
The second type, represented by the short red bars, consists of an isolated log where the parameter values remain constant.
This observation reveals that the disparate distribution of log groups enables a complementary parsing approach. For the log group rich in pattern variations, we can leverage the abundance of variable examples to perform lightweight statistical comparisons. 
For the isolated groups lacking such clues, we employ the semantic reasoning of LLMs. By tailoring the processing strategy to these distinctive categories, we can effectively bridge the trade-off between parsing cost and effectiveness.


\textbf{Definitions.} 
Based on the aforementioned observation, we formally categorize these groups into \textit{dense log groups} and \textit{sparse log groups}.
A \textit{dense log group} corresponds to the first type, defined as a collection of log messages belonging to the same template that exhibits dynamic patterns due to varying parameter values. 
Let a specific log $l(t,\vec{p})$ correspond to its log template $t$ with its log parameters $\vec{p}$. 
The dense log group for specific log template $\mathcal{G}_{D}(t)$ is defined as 
$$
\mathcal{G}_{D}(t) = \{ l(t,\vec{p}) | \vec{p} \in \mathbb{P}) \}
$$
where $\mathbb{P}$ is the parameter value space constituted by different parameter values.
Conversely, a \textit{sparse log group} exhibits a static appearance where parameter values remain constant. 
We define the sparse log group $\mathcal{G}_{S}(t)$ as
$$
\mathcal{G}_{S}(t) = \{ l(t,\vec{p})\}
$$
where the messages in $\mathcal{G}_{S}(t)$ share the identical parameters $\vec{p}$.

\begin{tcolorbox}[
    colback=pink!5!white,
    colframe=pink!75!black,
    fonttitle=\bfseries,
]
\small
\textbf{Lesson Learned:}
Log distributions exhibit two distinct patterns: \textit{statistically rich dense groups} and \textit{semantic-dependent sparse groups}. This division enables a hybrid strategy for log parsing that simultaneously achieves high effectiveness and low cost.

\end{tcolorbox}

\section{Methodology}
\tool is designed to exploit the strengths of both fast syntax parsers and highly effective semantic parsers. 
As shown in Figure~\ref{fig:overview}, \tool consists of a dynamic router and two parallel processors. 
First, the router processes log messages through a two-stage grouping phase to identify groups of logs with the same word skeleton (skeleton groups). 
Next, an anchor-based merging phase evaluates and merges similar skeleton groups. 
This step separates the logs into merged dense groups and unmerged sparse groups. 
The router then directs the dense groups to a statistical processor for efficient parsing, and the sparse groups to an LLM-based processor to identify complex variables.

\subsection{Dynamic Router}
Given log messages, the router aims to distinguish dense log groups from sparse ones. Crucially, this separation must be expedited to address latency concerns.
We achieve this by clustering similar logs through a two-stage grouping phase and a merging phase.

\subsubsection{Two-Stage Grouping Phase}
This phase aims to identify logs with high lexical similarity, as they typically share the same template.
We first abstract logs into masked skeletons by filtering out common variable tokens, which serve as the basic unit for all subsequent processing. Logs with identical skeletons form a skeleton group, and these groups are then further clustered by length.
This two-stage, multi-granular process allows it to capture various lexical characteristics of log messages.

\subsubsection*{Masked Skeleton-based Grouping.}
The first stage aims to construct fine-grained skeleton groups. 
Directly grouping raw logs often leads to excessive fragmentation due to the presence of dynamic variables. 
Therefore, we incorporate a preprocessing step using standard regular expressions, consistent with prior research~\cite{huang2025no,he2017drain}, to normalize the logs before parsing.
Specifically, we mask five variable structures and replace them with designated tokens. 
These structures include pure numbers (\texttt{<NUM>}), mixed strings with clear boundaries (\texttt{<CL>}), mixed strings without clear boundaries (\texttt{<UCL>}), capitalized short strings (\texttt{<BL>}), and single letters with delimiters (\texttt{<SL>}). 
After masking, we extract the remaining text as skeletons. 
We then group the logs based on these masked skeletons. 
For each skeleton group, its masked skeleton serves as a unique key. 
This key represents all individual logs within the group. By processing through the key, we eliminate redundant computations and accelerate all the subsequent steps. 

\begin{figure*}[tbp]
    \centering
    \includegraphics[width=0.7\linewidth]{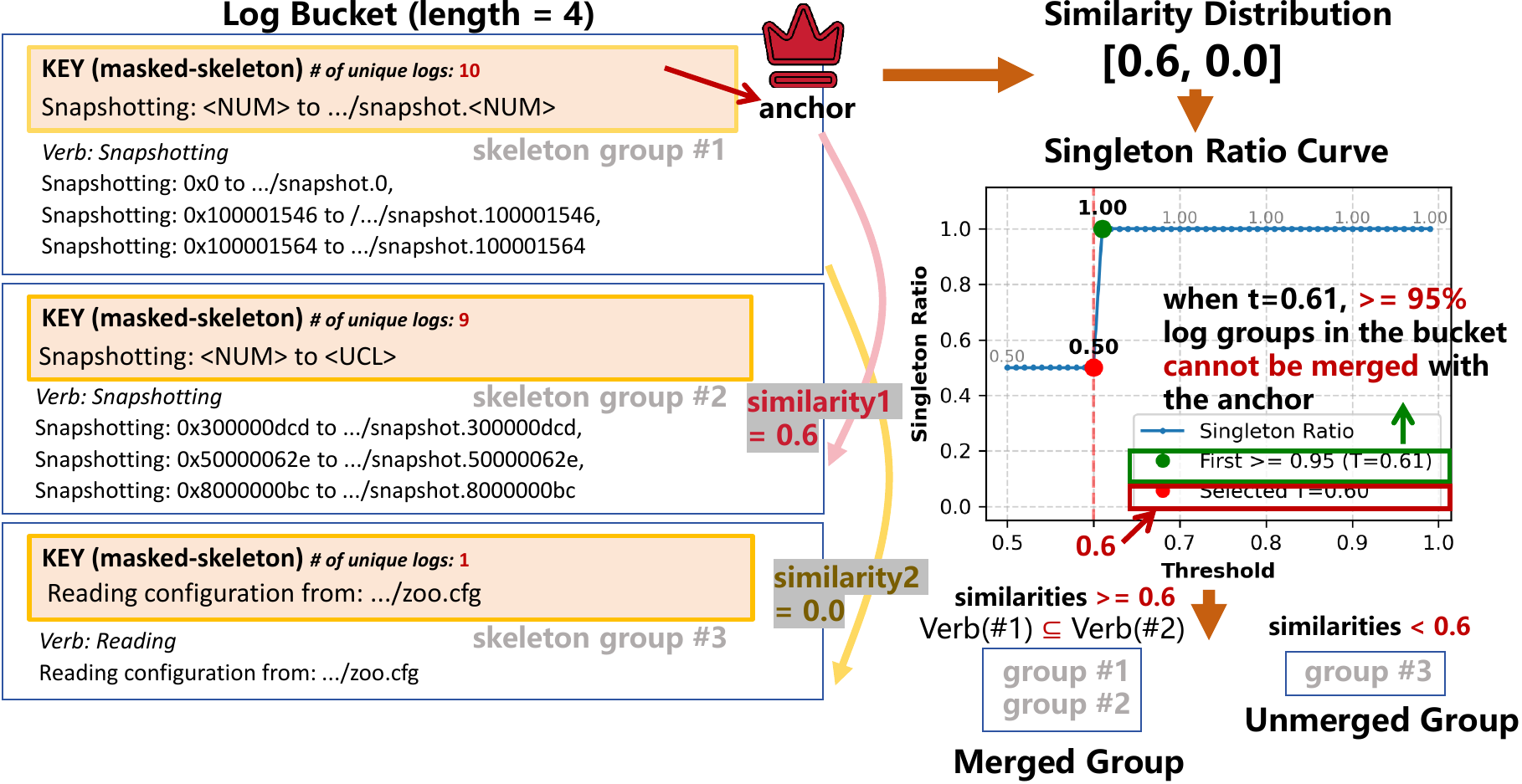}
    \caption{An example of the anchor-based merging phase of the Router.}
    \label{fig:exam:router}
\end{figure*}
\subsubsection*{{Length-based Grouping}}
The second stage applies a coarse grouping to the generated skeleton groups to prepare for efficient merging. 
While the strict first stage ensures high precision, it often splits logs of the same template into multiple skeleton groups, thus we need to combine these over-splitted groups for better approximation of the dense/sparse log groups. 
To limit the search space for potential matches, we aggregate the skeleton groups based on their length, since logs sharing a template typically have identical lengths~\cite{shima2016LenMa}. Specifically, we determine the length of each skeleton group by counting the whitespace-separated tokens in its key. Finally, we place groups of the same length into corresponding log buckets.


\begin{algorithm}[t]
\footnotesize
\caption{Anchor-based Merging (per log bucket)}
\label{alg:anchor-merge}
\begin{algorithmic}[1]
\REQUIRE Log bucket $B$
\ENSURE Set of dense log groups $\mathit{G}_{d}$, set of sparse log groups $\mathit{G}_{s}$
\STATE $\mathit{G}_{d} \leftarrow \emptyset$, $\mathit{G}_{s} \leftarrow \emptyset$
\IF{log length of $B$ is short \OR $|B|$ is small}
    \RETURN $\mathit{G}_{d} \leftarrow B$ \textcolor{gray}{\COMMENT{Directly route to the statistical processor}}
\ENDIF
\STATE $K \leftarrow \lfloor \alpha |B| \rfloor$ \textcolor{gray}{\COMMENT{Dynamically set top-$K$}}
\STATE Sort $B$ by unique log count descending
\WHILE{$B \neq \emptyset$ \AND $|\mathit{G}_{d}| < K$}
    \STATE $A \leftarrow B[0]$ \textcolor{gray}{\COMMENT{Recursively select Anchor}}
    \STATE Calculate similarities $\Sigma = \{ \text{PosJaccard}(A.key, x.key) \mid x \in B \}$
    \STATE $\tau \leftarrow \text{CalcThreshold}(\Sigma, \mathcal{P})$ \textcolor{gray}{\COMMENT{Set the threshold by $\mathcal{P}$ metric}}
    \STATE $Matched \leftarrow \{A\}$
    \FORALL{$x \in B \setminus \{A\}$}
        \IF{$\text{PosJaccard}(A.key, x.key) \geq \tau$ \AND $\text{Verbs}(A.key) \subseteq \text{Verbs}(x.key)$}
            \STATE $Matched \leftarrow Matched \cup \{x\}$
        \ENDIF
    \ENDFOR
    \STATE $\mathit{G}_{d} \leftarrow \mathit{G}_{d} \cup \{\text{Merge}(Matched)\}$
    \STATE $B \leftarrow B \setminus Matched$
\ENDWHILE
\STATE $\mathit{G}_{s} \leftarrow B$ \textcolor{gray}{\COMMENT{Add remaining groups as sparse groups}}
\RETURN $\mathit{G}_{d}$, $\mathit{G}_{s}$
\end{algorithmic}
\end{algorithm}
\subsubsection{Anchor-based Merging Phase}
This phase aims to selectively merge highly similar skeleton groups within the same bucket.
Groups that successfully merge together form dense log groups. Conversely, those that remain unmerged are classified as sparse log groups. 
The primary challenge lies in optimizing the matching efficiency. A naive exhaustive comparison among all groups leads to significant computational redundancy, especially as the number of unique skeletons scales. Moreover, employing a fixed similarity threshold lacks the flexibility needed to handle diverse log formats across different systems, often requiring extensive manual tuning.

To overcome these issues, we propose an anchor-based merging mechanism. 
Instead of pairwise comparisons, we select a few skeleton groups with the most unique logs as \textit{bucket anchors}. 
These \textit{anchors} generally represent the dominant portion of a dense log group.
We then rapidly compare the remaining groups only against these anchors. 
This avoids pairwise checks and greatly accelerates the process. 
Further, to ensure robustness, we evaluate similarity using a dynamic threshold.  


Algorithm~\ref{alg:anchor-merge} details the strategy.
The input log bucket $B$ consists of multiple skeleton groups.
We first assess the complexity of the bucket.
If the log length is short or the bucket contains few skeleton groups, we bypass the merging phase and route them directly to the statistical processor \textbf{(Lines 2--3)}. 
We assume these buckets have minimal fragmentation and stable structures. This avoids unnecessary computation and prevents excessive merging.
For complex buckets, we dynamically calculate a limit $K$ by setting $K$ as $\lfloor \alpha |B| \rfloor$, where $\alpha$ is a predefined proportion parameter \textbf{(Line 5)}. 
This adapts the limit to different bucket sizes.
Next, we sort the skeleton groups in descending order based on their unique log counts \textbf{(Line 6)}.
We then iteratively select the top-ranked group as the anchor $A$ and compute the similarity distribution $\Sigma$ against the remaining groups using Position-aware Jaccard similarity \textbf{(Lines 7--9)}.
Subsequently, we derive an adaptive threshold $\tau$ from $\Sigma$ to identify similar groups \textbf{(Lines 10--11)}.
We aggregate groups that satisfy both the similarity threshold $\tau$ and verb subset constraints.
We then merge them into a dense log group in $\mathit{G}_{d}$ \textbf{(Lines 12--17)}.
This iterative process continues until the bucket is empty or the number of identified dense log groups reaches the limit 
$K$ \textbf{(Lines 7--19)}.
Finally, any remaining unmerged groups are categorized as sparse groups $\mathit{G}_{s}$ \textbf{(Line 20)}.

\subsubsection*{Similarity Distribution \& Dynamic Similarity Threshold.}
As indicated in Lines 9--10 of Algorithm~\ref{alg:anchor-merge}, we employ a dynamic thresholding mechanism based on the distribution of similarity scores.

For a selected anchor $A$, we first compute the Position-aware Jaccard Similarity against all other skeleton groups in the bucket to form a similarity distribution $\Sigma$.
Unlike standard Jaccard Similarity~\cite{niwattanakul2013using}, which treats logs as unordered bags of words, Position-aware Jaccard Similarity considers both the token value and its position index, thereby penalizing structural mismatches.
To determine the optimal cut-off, we analyze the \textit{Singleton Ratio Curve} of $\Sigma$, which tracks the proportion of unmerged groups (singletons) as the similarity threshold $\tau$ increases from $0.5$ to $0.95$.
For instance, in Figure~\ref{fig:exam:router}, the anchor compares against two groups with similarity scores of $0.6$ and $0.0$. 
At a threshold of $0.6$, the first group merges while the second remains isolated. This yields one singleton out of two candidates, resulting in a ratio of $0.5~(1/2)$. Conversely, raising the threshold to $0.61$ excludes both groups. The singleton count then equals the total count, pushing the ratio to $1.0~(3/3)$.

We use the singleton ratio to determine the similarity threshold for merging skeleton groups with an anchor dynamically. 
An anchor represents a dominant part of a dense log group because it contains the most unique logs. 
Therefore, it should naturally merge with other members of the same dense group. 
However, an overly strict similarity threshold causes a high proportion of candidate groups to fail this merge and remain as singletons. 
To ensure that valid members are not excluded, we set a predefined strict limit~(i.e., $\mathcal{P}$) on the singleton ratio. Once the ratio reaches this limit, we revert to the immediately preceding similarity threshold. This prior value then serves as our final loose threshold.
This strategy ensures that we capture the maximal valid merging boundary just before the similarity constraint becomes overly restrictive, thus preventing over-merging while maintaining high precision.
\begin{figure}[tbp]
    \centering
    \includegraphics[width=0.8\linewidth]{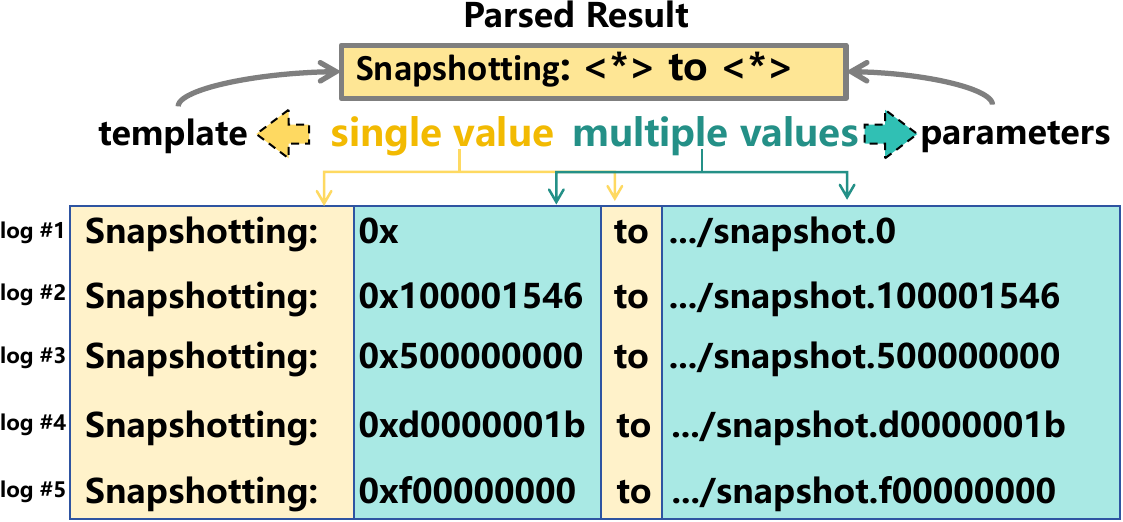}
    \caption{An example of the Statistical Processor.}
    \label{fig:exam:fast-track}
\end{figure}
\subsubsection*{Lightweight Verb Checking.}
We incorporate a lightweight verb constraint to ensure semantic consistency. 
Some skeleton groups exhibit high structural similarity but perform completely different actions. To distinguish them, we extract verbs from the keys of both the anchor and the target skeleton group. 
We merge a skeleton group with the anchor only if its verb set contains all the verbs of the anchor. This guarantees that they share the same core action. 
Moreover, this check operates directly on the static keys rather than raw logs. Therefore, it adds minimal computational overhead.

Figure~\ref{fig:exam:router} shows an example of the anchor-based merging phase. 
In this example, the bucket contains three skeleton groups~(i.e., Group \#1, \#2, and \#3) with the same length of $4$. 
The algorithm identifies Group \#1 as the anchor because it contains the most unique logs. 
It then calculates the similarity distribution, yielding $0.6$ for Group \#2 and 0.0 for Group \#3. According to the Singleton Ratio Curve, the ratio spikes to $1.0$~(with $\mathcal{P}=0.95$) at a threshold of $0.61$. 
Therefore, the algorithm selects the preceding stable value of $0.6$ as the adaptive threshold. 
Group \#2 is successfully merged with the anchor as it satisfies both this similarity threshold and the verb consistency check. Conversely, Group \#3 remains unmerged as a sparse group due to its distinct structure and action.

\subsection{Statistical Processor}
We design a straightforward yet efficient statistical processor to handle dense log groups identified by the router via a fast track. 
By leveraging robust statistical pattern extraction instead of expensive and potentially unstable LLM calls, it provides a highly efficient and consistent solution for dense log groups. 
To begin with, we analyze the value distribution at each position within the aligned log group to distinguish between unique values and multiple distinct values. We then directly mask the positions containing multiple distinct values as parameter positions. Finally, we apply a lightweight post-processing step following former practices~\cite{huang2025no,xiao2024free,jiang2024lilac} to refine the extracted templates and accommodate domain-specific formats.
Figure~\ref{fig:exam:fast-track} shows an example of the proposed statistical processor. 
As shown, the input consists of five aligned log entries. 
The processor identifies the tokens \texttt{Snapshotting:} and \texttt{to} as single values shared by all logs. These are retained as static template parts. 
Conversely, the hexadecimal strings (e.g., \texttt{0x0} and \texttt{0x100001546}) and file paths~(e.g., \texttt{.../snapshot.0} and \texttt{.../snapshot.100001546}) exhibit multiple distinct values. These variable positions are replaced by wildcards~(e.g., \texttt{<*>}). Consequently, the processor outputs the template \texttt{Snapshotting: <*> to <*>} and extracts the corresponding parameters.

\begin{figure}[t]
    \centering
    \includegraphics[width=0.6\linewidth]{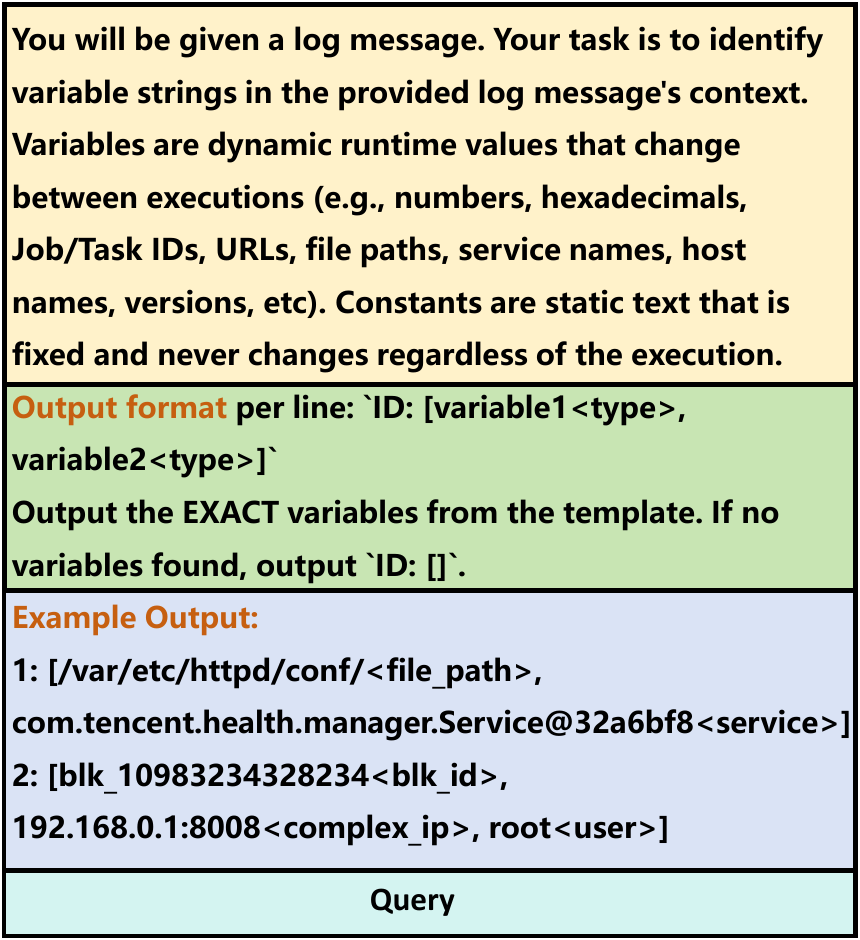}
    \caption{The prompt of the LLM-based processor.}
    \label{fig:prompt-template}
\end{figure}
\subsection{LLM-based Processor}
We propose the LLM-based processor to handle sparse log groups that lack sufficient signals for statistical analysis.
Our task design for the LLM focuses on variable identification rather than full template generation to minimize hallucination risks.
Specifically, we instruct the model to identify dynamic tokens and return them as a list of exact variable strings.
To ensure robustness, we process the LLM's output by verifying that the identified variables are indeed present in the original log message.
We then mask these validated variable substrings with a placeholder (e.g., \texttt{<*>}) to generate the final template.
If the LLM fails to identify any variables or produces an invalid format, we roll back to the original log to prevent erroneous over-masking.
Figure~\ref{fig:prompt-template} shows the prompt template. 
The prompt comprises a task description, output constraints, and fixed-output examples following former works~\cite{huang2025no}. 
This structured design guides the model to produce stable and accurate results.

\subsection{Parallel Processing}
To address the scalability challenges in large-scale log parsing, we implement a two-level parallelization strategy within \tool.
First, we exploit data parallelism during the routing phase by partitioning log buckets based on their token lengths.
These buckets are processed independently across multiple CPU cores, allowing for concurrent anchor identification and similarity computation without cross-bucket dependencies.
Second, for the LLM-based processor, we adopt asynchronous I/O to handle network latency.
By batching API requests and utilizing non-blocking calls, we maximize throughput and ensure that the system remains responsive even when interacting with external model services.
This hybrid approach effectively decouples computational tasks from I/O-bound operations, significantly reducing the overall parsing time.

\section{Evaluation}
We conduct evaluations to answer the following research questions (RQs):

\textbf{RQ1: How does \tool perform in log parsing?}

\textbf{RQ2: How do different components contribute to \tool?}

\textbf{RQ3: How sensitive is \tool to the parameter settings?}

\textbf{RQ4: How robust is \tool with
different backbone LLMs?}

\subsection{Dataset}
We evaluate \tool on 14 publicly available log datasets~\cite{zhu2023loghub,jiang2024large}, aligning with prior studies~\cite{jiang2024lilac,huang2025no}.
These datasets originate from a diverse array of systems, including distributed systems, supercomputers, operating systems, and server applications.
This collection covers a wide range of log scales and complexities, amounting to millions of log messages and thousands of unique templates, as summarized in Table~\ref{tab:dataset}.
Such diversity provides a comprehensive benchmark for evaluating the parsing effectiveness and cost of \tool across different log datasets.

\begin{table}[!tbp] 
\footnotesize
\caption{The statistics of the 14 public log datasets.}
\vspace{-0.1in}
\label{tab:dataset}
\begin{tabular}{cccc}
\toprule
\textbf{Systems} & \textbf{Datasets} & \textbf{\# Logs} & \textbf{\# Templates} \\
\toprule
Mobile systems 
    & HealthApp & 212,394 & 156 \\
\addlinespace
\multirow{2}{*}{Server applications} 
    & OpenSSH & 638,946 & 38 \\
    & Apache & 51,977 & 29 \\
\addlinespace
Standalone software 
    & Proxifier & 21,320 & 11 \\
\addlinespace
\multirow{3}{*}{Supercomputers} 
    & Thunderbird & 16,601,745 & 1,241 \\
    & BGL & 4,631,261 & 320 \\
    & HPC & 429,987 & 74 \\
\addlinespace
\multirow{2}{*}{Operating systems}
    & Mac & 100,314 & 626 \\
    & Linux & 23,921 & 338 \\
\addlinespace
\multirow{5}{*}{Distributed systems} 
    & Spark & 16,075,117 & 236 \\
    & HDFS & 11,167,740 & 46 \\
    & OpenStack & 207,632 & 48 \\
    & Hadoop & 179,993 & 236 \\
    & Zookeeper & 74,273 & 89 \\
\bottomrule
\end{tabular}
\end{table}

\subsection{Baselines}
Following prior work~\cite{jiang2024lilac,huang2025no}, we compare \tool with six state-of-the-art log parsers, comprising three syntax-based approaches and three LLM-based semantic approaches.
For the {syntax-based} parsers, we select {Drain}~\cite{he2017drain}, {AEL}~\cite{jiang2008abstracting}, and {Brain}~\cite{yu2023brain}. These methods are widely recognized for their efficiency and strong performance among traditional syntax-based parsers.
For the {semantic-based} approaches, we choose {LILAC}~\cite{jiang2024lilac}, {LogBatcher}~\cite{xiao2024free}, and {LUNAR}~\cite{huang2025no}.
Since our method operates in an unsupervised manner without labeled data, we adopt the label-free variant of LILAC\footnote{Unless otherwise specified, we use LILAC to denote the label-free variant (LILAC w/o ICL) in the remainder of this paper.} (i.e., LILAC w/o ICL) for a fair comparison, as the original version relies on in-context learning with labeled examples.
All baselines are evaluated using their official open-source implementations with default parameters to ensure reproducibility.

\subsection{Metrics}
Following prior practices~\cite{jiang2024lilac,huang2025no}, we evaluate \tool from two primary aspects: parsing effectiveness and parsing cost.

\subsubsection{Parsing Effectiveness}
We use four standard metrics to assess parsing quality at both the message and template levels in alignment with existing works~\cite{huang2025no,jiang2024lilac,jiang2024large}.

$\bullet$ \textbf{Grouping Accuracy (GA)} calculates the ratio of log messages assigned exactly to their correct ground truth clusters.

$\bullet$ \textbf{Parsing Accuracy (PA)} measures the ratio of messages where every token is correctly classified as a constant or a variable.

$\bullet$ \textbf{F1 score of Grouping Accuracy (FGA)} evaluates grouping performance at the template level. It calculates the harmonic mean of grouping precision and recall to address class imbalance.

$\bullet$ \textbf{F1 score of Template Accuracy (FTA)} evaluates both grouping correctness and text accuracy. It calculates the harmonic mean of template precision and recall. A template is correct only if its grouping and text perfectly match the ground truth.

\subsubsection{Parsing Cost}
To assess the efficiency and economic feasibility of our approach, we employ three cost-related metrics.

$\bullet$ \textbf{Parsing time} records the total end-to-end execution time. This includes preprocessing, model inference, and post-processing.

$\bullet$ \textbf{Token consumption} measures the total number of tokens processed by the LLM. It is a direct proxy for the economic cost.

$\bullet$ \textbf{LLM invocations} counts the total number of API calls made to the LLM service. Reducing these calls minimizes network overhead and prevents rate limits.

\subsection{Environment and Implementation}
All experiments are conducted on a MacBook Pro equipped with an Apple M4 Pro processor with 12 cores and 24GB of unified memory, running macOS Sequoia.
We employ GPT-5.2-2025-12-11~\cite{gpt} as the default LLM accessed via the standard OpenAI API interface~\cite{openai-api}.
To ensure reproducibility and eliminate randomness, the temperature parameter is set to 0. 
We use  Spacy~\cite{spacy} with NLTK WordNet~\cite{nltk,bird2009natural} for lightweight verb validation.
Regarding the hyperparameters, we configure the default dynamic similarity threshold using the P95 criterion~(i.e., $\mathcal{P}=0.95$) based on the singleton ratio curve, and set the top-$k$ anchors dynamically to half of the log bucket length (i.e., $k = \lfloor |B|/2 \rfloor$ with $\alpha = 1/2$), which is further analyzed in our sensitivity study.
For the parallel version of \tool, we utilize 8 CPU cores to process log buckets concurrently.
In contrast, the non-parallel version operates sequentially on a single core to provide a baseline for efficiency comparison.

\begin{table*}[tbp] 
\footnotesize
\centering \caption{Effectiveness comparison with SOTA parsers (best: bold; second-best: underlined; `-': 24h timeout).} \label{effectiveness_result_table} \begin{adjustbox}{max width = \textwidth} \begin{tabular}{c|c|ccccccccccccccc} 
\toprule 
\textbf{Method} & \textbf{Metric} & Apache  & BGL & Hadoop & HDFS & HealthApp & HPC & Linux & Mac & OpenSSH & OpenStack & Proxifier & Spark & Thunderbird & Zookeeper & \textbf{Average} \\ 
\toprule 
\rowcolor{gray!20}\multicolumn{17}{c}{\textbf{Syntax-based Log Parsers}} \\ 

\midrule  
\multirow{4}{*}{\textbf{Drain}}  
& GA & \textbf{1.000} & 0.919 & 0.921 & \underline{0.999} & 0.862 & 0.793 & 0.686 & 0.761 & 0.707 & 0.752 & 0.692 & 0.888 & \underline{0.831} & \underline{0.994} & 0.843\\  
& PA & 0.727 & 0.456 & 0.546 & 0.569 & 0.312 & 0.721 & 0.112 & 0.381 & 0.586 & 0.020 & \underline{0.688} & 0.559 & 0.219 & 0.844 & 0.481\\  
& FGA & \textbf{1.000} & 0.624 & 0.785 & \underline{0.935} & 0.010 & 0.309 & 0.778 & 0.230 & 0.872 & 0.007 & 0.206 & \underline{0.861} & 0.237 & \underline{0.904} & 0.554\\  
& FTA & 0.517 & 0.204 & 0.394 & 0.478 & 0.004 & 0.147 & 0.262 & 0.070 & 0.487 & 0.002 & 0.176 & 0.448 & 0.072 & 0.627 & 0.278\\

\toprule 
\multirow{4}{*}{\textbf{AEL}} 
& GA & \textbf{1.000} & 0.915 & 0.823 & \underline{0.999} & 0.725 & 0.748 & \underline{0.917} & 0.797 & 0.705 & 0.743 & 0.974 & - & 0.786 & \textbf{0.996} & 0.856\\  
& PA & 0.727 & 0.455 & 0.540 & 0.569 & 0.311 & 0.741 & 0.093 & 0.269 & 0.364 & 0.020 & 0.677 & - & 0.166 & 0.842 & 0.444\\  
& FGA & \textbf{1.000} & 0.587 & 0.117 & 0.764 & 0.008 & 0.201 & 0.812 & 0.793 & 0.689 & 0.682 & 0.667 & - & 0.116 & 0.788 & 0.556\\  
& FTA & 0.517 & 0.176 & 0.060 & 0.494 & 0.003 & 0.133 & 0.259 & 0.211 & 0.333 & 0.141 & 0.417 & - & 0.036 & 0.475 & 0.250\\

\toprule  
\multirow{4}{*}{\textbf{Brain}}
& GA & 0.997 & 0.940 & 0.503 & 0.960 & 0.865 & 0.800 & 0.790 & 0.808 & 0.663 & \textbf{1.000} & 0.521 & 0.838 & 0.792 & {0.993} & 0.819\\  
& PA & 0.262 & 0.435 & 0.123 & 0.722 & 0.308 & 0.662 & 0.067 & 0.324 & 0.261 & 0.149 & 0.687 & 0.556 & 0.199 & 0.818 & 0.398\\  
& FGA & 0.933 & 0.756 & 0.511 & 0.759 & 0.864 & 0.400 & 0.749 & 0.737 & 0.759 & \textbf{1.000} & 0.737 & 0.207 & 0.748 & 0.798 & 0.711\\  
& FTA & 0.433 & 0.179 & 0.188 & 0.437 & 0.411 & 0.184 & 0.230 & 0.277 & 0.299 & 0.312 & 0.421 & 0.086 & 0.260 & 0.546 & 0.305\\

\midrule 
\rowcolor{gray!20}\multicolumn{17}{c}{\textbf{LLM-based Log Parsers}} \\ 

\midrule  
\multirow{4}{*}{\textbf{LILAC}} 
& GA & \textbf{1.000} & 0.880 & 0.915 & \textbf{1.000} & \textbf{0.999} & \underline{0.871} & 0.641 & 0.750 & 0.672 & 0.554 & 0.059 & 0.892 & 0.820 & {0.993} & 0.789\\  
& PA & \underline{0.992} & \underline{0.940} & \underline{0.811} & 0.569 & 0.582 & \textbf{0.979} & 0.642 & 0.532 & 0.365 & 0.496 & 0.189 & \underline{0.874} & \underline{0.580} & 0.394 & 0.639\\  
& FGA & \textbf{1.000} & \underline{0.886} & \textbf{0.913} & 0.658 & \textbf{0.978} & \textbf{0.922} & 0.739 & 0.815 & 0.708 & \underline{0.949} & 0.062 & \textbf{0.863} & 0.811 & 0.900 & 0.800\\  
& FTA & 0.816 & \textbf{0.784} & \textbf{0.749} & 0.411 & \underline{0.830} & \textbf{0.828} & 0.567 & \underline{0.539} & 0.503 & \textbf{0.841} & 0.211 & \textbf{0.695} & 0.543 & 0.688 & 0.643\\

\toprule 
\multirow{4}{*}{\textbf{LogBatcher}} 
& GA & 0.997 & \textbf{0.947} & \underline{0.929} & \underline{0.999} & 0.942 & 0.864 & 0.832 & \underline{0.892} & 0.749 & \textbf{1.000} & \textbf{1.000} & \textbf{0.973} & 0.684 & {0.993} & \underline{0.914}\\ 
& PA & 0.968 & 0.909 & 0.736 & 0.947 & 0.767 & 0.883 & 0.627 & 0.603 & 0.702 & \textbf{0.933} & \textbf{1.000} & 0.684 & 0.420 & 0.827 & 0.786\\ 
& FGA & 0.918 & \textbf{0.903} & 0.870 & \textbf{0.968} & \underline{0.950} & \underline{0.859} & \underline{0.863} & \underline{0.845} & \underline{0.909} & \textbf{1.000} & \textbf{1.000} & 0.829 & \underline{0.829} & \textbf{0.978} & \textbf{0.909}\\  
& FTA & 0.721 & \underline{0.772} & 0.662 & 0.731 & 0.785 & \underline{0.817} & \underline{0.667} & 0.486 & 0.701 & 0.771 & \textbf{1.000} & 0.596 & \underline{0.551} & \textbf{0.844} & \underline{0.722}\\

\toprule 
\multirow{4}{*}{\textbf{LUNAR}} 
& GA & \underline{0.999} & \underline{0.946} & \textbf{0.938} & 0.916 & \underline{0.992} & \textbf{0.907} & 0.706 & 0.874 & \underline{0.761} & \underline{0.968} & 0.669 & \underline{0.906} & \textbf{0.860} & {0.993} & 0.888\\  
& PA & 0.914 & \textbf{0.970} & \textbf{0.897} & \underline{0.960} & \textbf{0.944} & 0.922 & \underline{0.742} & \underline{0.619} & \underline{0.703} & 0.892 & 0.680 & \textbf{0.947} & \textbf{0.584} & \underline{0.850} & \underline{0.830}\\ 
& FGA & \underline{0.978} & 0.773 & 0.692 & 0.892 & 0.932 & 0.753 & 0.827 & 0.768 & \textbf{0.911} & 0.566 & 0.775 & 0.853 & 0.827 & 0.885 & 0.816\\  
& FTA & 0.728 & 0.721 & 0.562 & \textbf{0.904} & 0.731 & 0.814 & 0.608 & 0.506 & \underline{0.911} & 0.508 & 0.864 & 0.658 & 0.533 & 0.792 & 0.703\\

\midrule 
\rowcolor{gray!20}\multicolumn{17}{c}{\textbf{Our Proposed Method (Hybrid)}} \\ \midrule  

\multirow{4}{*}{\textbf{\tool}} 
& GA & 0.997 & 0.944 & 0.919 & \textbf{1.000} & 0.984 & 0.861 & \textbf{0.940} & \textbf{0.916} & \textbf{0.780} & 0.960 & \underline{0.989} & 0.889 & 0.815 & 0.988 & \textbf{0.927}\\  
& PA & \textbf{0.997} & 0.937 & 0.768 & \textbf{1.000} & \underline{0.928} & \underline{0.932} & \textbf{0.857} & \textbf{0.632} & \textbf{1.000} & \underline{0.894} & \textbf{1.000} & 0.855 & 0.451 & \textbf{0.993} & \textbf{0.875}\\  
& FGA & 0.944 & 0.836 & \underline{0.892} & 0.894 & 0.936 & 0.686 & \textbf{0.867} & \textbf{0.867} & \textbf{0.911} & 0.936 & \underline{0.870} & 0.842 & \textbf{0.862} & 0.882 & \underline{0.873}\\  
& FTA & \textbf{0.876} & 0.742 & \underline{0.692} & \underline{0.830} & \textbf{0.861} & 0.705 & \textbf{0.747} & \textbf{0.541} & \textbf{0.962} & \underline{0.830} & \underline{0.957} & \underline{0.668} & \textbf{0.592} & \underline{0.816} & \textbf{0.773}\\
\cline{1-17}   \end{tabular} \end{adjustbox} \end{table*}
\subsection{RQ1: Comparison Study}
\begin{figure}[tbp]
  \centering
  \begin{minipage}[b]{0.48\columnwidth}
    \centering
    \includegraphics[width=\textwidth]{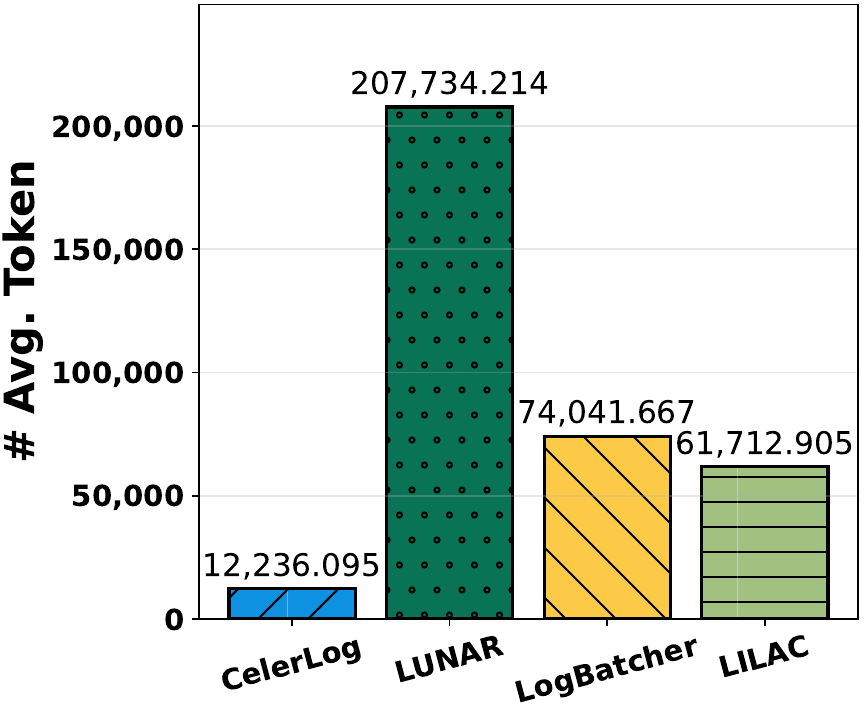}
    \caption{Average LLM token consumption of LLM-based log parsers and \tool. \tool achieves the lowest token consumption.}
    \label{fig:rq1-1}
  \end{minipage}
  \hfill 
  \begin{minipage}[b]{0.48\columnwidth}
    \centering
    \includegraphics[width=\textwidth]{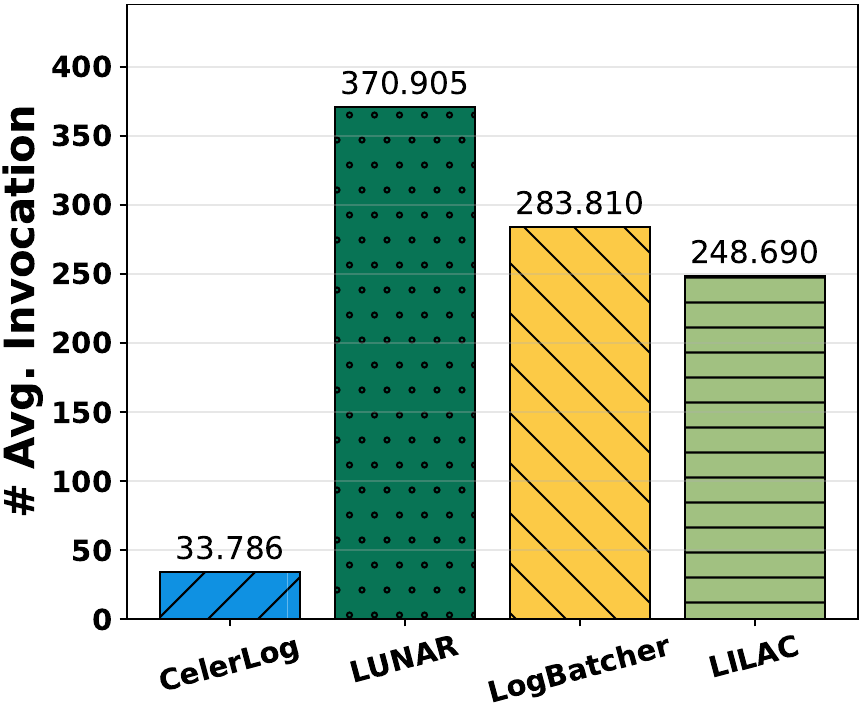}
    \caption{Average LLM invocations for LLM-based log parsers and \tool. \tool achieves the lowest invocations.}
    \label{fig:rq1-2}
  \end{minipage}
\end{figure}
\begin{figure}[t]
    \centering
    \includegraphics[width=\linewidth]{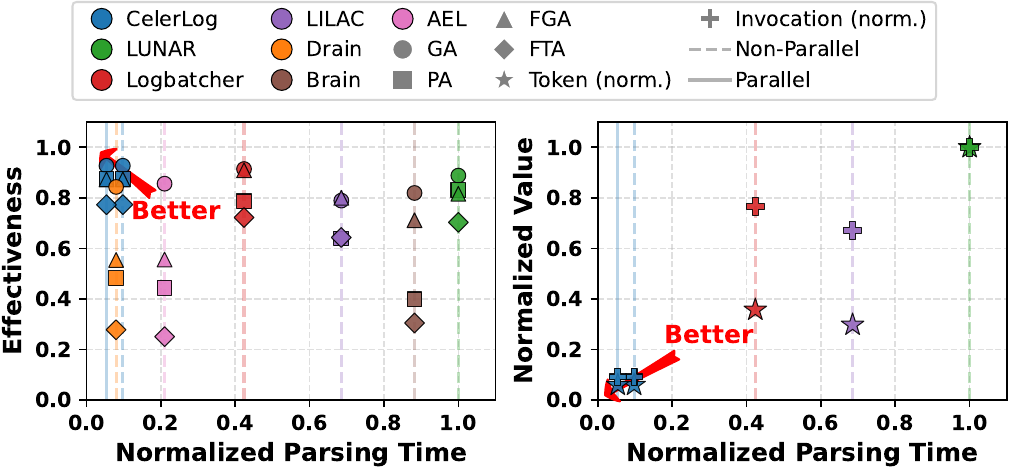}
    \caption{Performance and cost trade-offs of log parsers. Left: Top-left is optimal (higher effectiveness, lower time). Right: Bottom-left is optimal (lower resources, lower time). Values are scaled to $[0, 1]$ relative to the maximum observed.}
    \label{rq1-versus}
\end{figure}
\begin{table*}[tbp]
    \small
    \caption{Parsing time (s) comparison with SOTA log parsers (best: bold; second-best: underlined; `-': 24h timeout).}
    \vspace{-0.1in}
    \label{tab:rq1-cs}
    \centering
    \begin{tabular}{l c c c c cc cc}
    \toprule
     \rowcolor{gray!20}\cellcolor{white}\multirow{3}{*}{Dataset} &  \multicolumn{3}{|c}{Syntax-based} & \multicolumn{3}{|c}{LLM-based} & \multicolumn{2}{|c}{Proposed}\\
    &
    {\multirow{2}{*}{\textbf{Drain}}} & 
    {\multirow{2}{*}{\textbf{AEL}}} & 
    {\multirow{2}{*}{\textbf{Brain}}} & 
    {\multirow{2}{*}{\textbf{LILAC}}} & 
    {\multirow{2}{*}{\textbf{LogBatcher}}} & 
    {\multirow{2}{*}{\textbf{LUNAR}}} &
    \multicolumn{2}{c}{\textbf{\tool}} \\
     & & & & & & & \textit{non-parallel} & \textit{parallel} \\
    \toprule

        Apache & \underline{1.094} & 1.971 & \textbf{1.079} & 61.125 & 51.991 & 56.458 & 8.507 & 3.932\\ 
 
        BGL & 98.422 & \underline{58.906} & 144.249 & 1834.448 & 573.258 & 1519.758 & 97.040 & \textbf{28.572}\\ 

        Hadoop & \textbf{4.304} & 28.746 & \underline{8.020} & 546.029 & 417.418 & 946.542 & 69.583 & 10.740\\ 
        
        HDFS & \underline{285.788} & 658.768  & 1325.486 & 368.709 & \textbf{118.989} & 1448.672 & 305.582 & 358.084\\ 
        
        HealthApp & \textbf{3.747} & 141.797 & 28.364 & 267.664 & 245.609 & 261.333 & 43.880 & \underline{8.583}\\ 
        
        HPC & 8.632 & 9.739  & \underline{7.486} & 175.825 & 144.696 & 288.611 & 18.480 & \textbf{5.542}\\ 
        
        Linux & \textbf{0.600} & 1.833 & \underline{0.641} & 659.839 & 620.941 & 545.985 & 143.284 & 18.656\\ 
        
        Mac & \underline{3.271} & 5.306 & \textbf{2.938} & 1259.866 & 1243.351 & 2354.536 & 174.201 & 22.673 \\ 
        
        OpenSSH & \underline{15.976} & 114.593 & 22.121 & 78.295 & 68.868 & 139.490 & 16.763 & \textbf{12.063} \\ 
        
        OpenStack & 20.504 & 11.794 & \textbf{8.253} & 121.632 & 81.821 & 391.994 & 18.952 & \underline{9.147}\\
        
        Proxifier & 0.769 & 1.094 & \underline{0.537} & 29.232 & 20.811 & 24.474 & \textbf{0.577} & 0.588 \\
        
        Spark  & 296.517 & - & 3584.402 & 892.430 & 536.702 & 2405.993 & \underline{274.546} & \textbf{259.544} \\
        
        Thunderbird & 513.338 & 2037.152 & 8722.304 & 4339.079 & 2414.931 & 5233.850 & \underline{345.401} & \textbf{104.551} \\ 
        
        Zookeeper & \textbf{1.571} & \underline{1.658} & 51.395 & 169.257 & 139.679 & 148.625 & 22.613 & 4.165 \\ 
    \bottomrule
    Average & \underline{89.610} & > 236.412  & 993.377 & 771.674 & 477.076 & 1126.166 & 109.958 & \textbf{60.489}\\

    Speedup & 1.5$\times$& > 3.9$\times$ & 16.4$\times$ & 12.8$\times$ & 7.9$\times$ & 18.6$\times$ & 1.8$\times$ & --\\
    \bottomrule
    \end{tabular}

\end{table*}

We evaluate \tool’s performance against state-of-the-art baselines in terms of parsing effectiveness and cost. On average, 98.7\% of logs are identified as dense groups, while only 1.3\% fall into sparse groups.
Table~\ref{effectiveness_result_table} presents the comparison of \tool against syntax-based parsers (Drain, AEL, Brain) and LLM-based parsers (LILAC, LogBatcher, LUNAR). 
Syntax-based methods generally struggle with complex log patterns, as evidenced by Drain's lower PA of 0.481. While LLM-based approaches leverage semantic understanding to improve accuracy, they often lack consistency in fine-grained metrics. 
\tool outperforms all baselines across most datasets, achieving the highest GA of 0.927 and PA of 0.875. 
Notably, \tool surpasses the best-performing LLM-based baseline, LogBatcher, by a significant margin in PA (0.875 vs. 0.786). 

Beyond accuracy, parsing efficiency is critical for practical deployment. Table~\ref{tab:rq1-cs} details the parsing time for all methods. Pure LLM-based methods suffer from high latency due to massive model inference, with LUNAR and LILAC requiring 1126.166 seconds and 771.674 seconds on average, respectively. 
In contrast, \tool (parallel) drastically reduces the average parsing time to 60.489 seconds, achieving an 18.6$\times$ speedup over LUNAR. Remarkably, \tool is even (1.5$\times$) faster than the widely-deployed parser Drain (89.610 seconds).
In terms of economic cost, Figure~\ref{fig:rq1-1} and Figure~\ref{fig:rq1-2} highlight the resource consumption. Existing LLM-based parsers incur heavy token usage. 
For instance, LUNAR consumes approximately 207,734 tokens on average. \tool reduces this consumption to merely 12,236 tokens, representing a huge reduction. Similarly, Figure~\ref{fig:rq1-2} shows that \tool requires only 33.786 LLM invocations on average, whereas competing methods like LUNAR require over 370 invocations. 

\textbf{Effectiveness vs. Cost Trade-off.} Figure~\ref{rq1-versus} visualizes the balance between parsing effectiveness and parsing cost. 
The left plot demonstrates that \tool resides in the top-left region, indicating superior parsing effectiveness with minimal time overhead. The right plot confirms that \tool achieves the lowest normalized cost while maintaining high parsing speed. Consequently, \tool establishes a new SOTA Pareto frontier, offering a robust solution that is both highly accurate and cost-effective for large-scale log parsing.

\begin{tcolorbox}
[
    colback=gray!5!white,
    colframe=gray!75!black,
    fonttitle=\bfseries,
    sharp corners
]
\small
\vspace{-0.1in}
\textbf{Answer to RQ1:}
\tool not only achieves leading performance and runs tenfold faster ($7.9\times$--$18.6\times$) than baselines, but also significantly reduces costs by decreasing token consumption by $80.2\%$--$94.1\%$ and LLM invocations by $86.4\%$--$90.9\%$.

\vspace{-0.1in}
\end{tcolorbox}

\begin{table*}[tbp]
\centering
\footnotesize
\caption{Ablation study of components of \tool. We use the non-parallel version to ensure clearer parsing time comparisons.}
\label{tab:ablation_study}
\begin{tabular}{l|cccc|ccc}
\toprule

\rowcolor{gray!20}\cellcolor{white}\multirow{2}{*}{\tool} & \multicolumn{4}{c|}{\textit{Effectiveness}} & \multicolumn{3}{c}{\textit{Cost}} \\

& GA & PA & FGA & FTA & Time (s) & \# Token & \# Invocation\\

\midrule

Full & 0.927 & 0.875 & 0.873 & 0.773 & 109.958 & 12,236.095 & 33.786 \\
w/o router & 0.696 ($\downarrow$24.9\%) & 0.781($\downarrow$10.7\%) & 0.676 ($\downarrow$22.6\%) & 0.637 ($\downarrow$17.6\%) & 325.958 ($\uparrow$196.4\%)& 52,814.071 ($\uparrow$331.6\%) & 148.357 ($\uparrow$339.1\%) \\

w/o statistical proc. & 0.916 ($\downarrow$1.2\%) & 0.622 ($\downarrow$28.9\%) & 0.873 (--) & 0.604 ($\downarrow$21.9\%) & 270.748 ($\uparrow$146.2\%) & 43,696.071 ($\uparrow$257.1\%) & 181.143 ($\uparrow$436.2\%)\\

w/o LLM proc. & 0.908 ($\downarrow$2.1\%) & 0.815 ($\downarrow$6.9\%) & 0.680 ($\downarrow$22.1\%) & 0.473 ($\downarrow$38.8\%) & 52.308 ($\downarrow$52.4\%) & - & - \\

\bottomrule
\end{tabular}
\end{table*}

\subsection{RQ2: Ablation Study}
To answer the question, we conduct an ablation study by comparing the full model with three variants: 
(1) \textbf{w/o router}, which randomly assigns logs to either the statistic or LLM processor instead of using the proposed routing mechanism; 
(2) \textbf{w/o statistical processor}, which selects the top three most frequent log messages from a dense group for LLM querying, replacing the statistical processor; 
and (3) \textbf{w/o LLM processor}, which removes the LLM component entirely.
Table~\ref{tab:ablation_study} presents the parsing effectiveness and cost comparison for these variants.

As the results indicate, the router is critical for balancing efficiency and effectiveness. 
Removing the router causes a severe degradation in parsing effectiveness, with GA dropping by 24.9\% and PA by 10.7\%. More importantly, the operational cost skyrockets: token consumption and LLM invocations increase by 331.6\% and 339.1\%, respectively. This indicates that the router effectively identifies and routes dense and sparse log groups, ensuring that expensive LLM resources are reserved only for sparse log groups.
Similarly, the statistical processor plays a crucial role in ensuring stability and efficiency when handling dense log groups. The \textit{w/o statistical processor} variant results in a 28.9\% decrease in PA and a 257.1\% surge in token usage. This significant drop in effectiveness suggests that relying solely on the LLM for high-frequency logs introduces instability, primarily due to LLM hallucinations where parameters are often misidentified or over-extracted. By leveraging statistical patterns for dense groups, \tool mitigates these generative errors and achieves robust parsing results with substantially lower computational cost.
As for the LLM processor, it provides the critical semantic understanding required for high-precision parsing. While the \textit{w/o LLM processor} variant reduces parsing time by 52.4\%, it causes a substantial degradation in fine-grained metrics, with FTA dropping by 38.8\% and FGA by 22.1\%. This result confirms that traditional syntax-based approaches are insufficient for capturing the semantic complexity of modern system logs, highlighting the indispensable role of the LLM in achieving the superior accuracy of \tool.


\begin{tcolorbox}
[
    colback=gray!5!white,
    colframe=gray!75!black,
    fonttitle=\bfseries,
    sharp corners
]
\small
\vspace{-0.1in}
\textbf{Answer to RQ2:}
Every component in \tool is essential, as the router maximizes efficiency, the statistical processor delivers stable templates, and the LLM-based processor provides semantic awareness for sparse log groups.
\vspace{-0.1in}
\end{tcolorbox}

\subsection{RQ3: Sensitivity Analysis}
\subsubsection{Dynamic Similarity Threshold}
\begin{figure}[t]
    \centering
    \includegraphics[width=\linewidth]{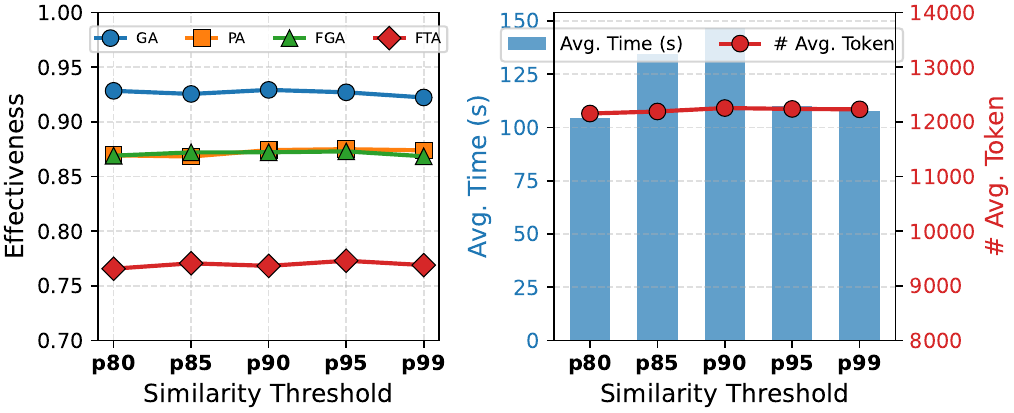}
    \caption{The parsing results under different similarity thresholds. The performance and cost remain stable across various percentile settings, demonstrating the robustness of our dynamic thresholding mechanism.}
    \label{fig:similarity-threshold}
\end{figure}

Figure~\ref{fig:similarity-threshold} presents the parsing effectiveness and cost under different percentile settings for the dynamic similarity threshold (from P80 to P99).
The results demonstrate that \tool is highly insensitive to this hyperparameter.
Across the entire range, both effectiveness metrics and efficiency metrics remain remarkably stable with minimal fluctuations.
This stability confirms the robustness of our singleton ratio curve-based approach, which adaptively determines the optimal cut-off for each log bucket rather than relying on a rigid global threshold.
Consequently, we select P95 as the default setting to ensure a high safety margin against over-merging while maintaining excellent clustering performance.

\subsubsection{Top-$k$ Selection}
\begin{figure}[t]
    \centering
    \includegraphics[width=\linewidth]{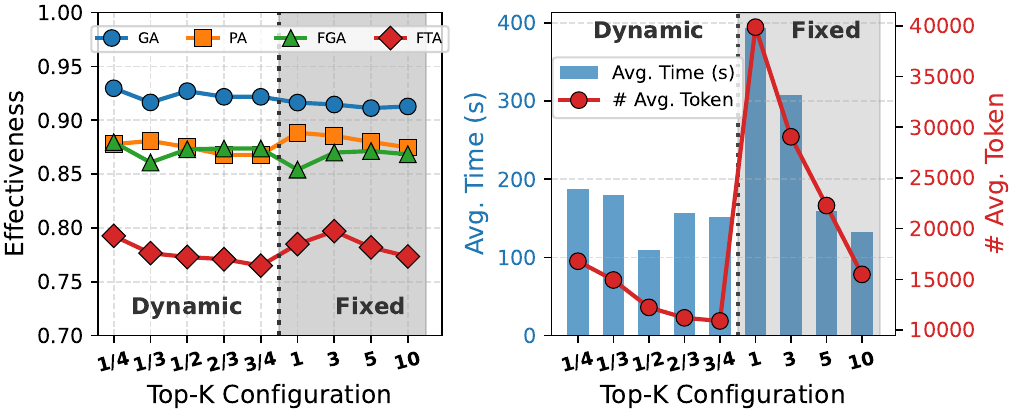}
    \caption{The parsing results under different top-$k$ settings. The dynamic settings (fractions) exhibit significantly better stability and cost-effectiveness compared to the fixed settings (integers), with the $1/2$ ratio offering the optimal trade-off.}
    \label{fig:anchor-number}
\end{figure}

Figure~\ref{fig:anchor-number} illustrates the impact of different top-$k$ anchor selection strategies on parsing effectiveness and cost.
We compare two settings: (1) \textit{Fixed setting}, where $k$ is a constant integer (e.g., $k=1, 3, 5, 10$); and (2) \textit{Dynamic setting}, where $k$ is a fraction of the log bucket capcity (e.g., $1/4, 1/3, 1/2, 2/3, 3/4$).

\textbf{Sensitivity of Fixed Settings.} As shown in the right part of Figure~\ref{fig:anchor-number}, the fixed setting exhibits high sensitivity and volatility.
A small fixed $k$ (e.g., $k=1$) fails to capture sufficient anchors for complex buckets, leading to a surge in sparse log groups.
This forces the downstream LLM processor to handle a significantly larger volume of logs, resulting in a dramatic spike in parsing time and token consumption (e.g., average time > 350s for $k=1$).
Conversely, a large fixed $k$ increases the risk of selecting low-quality anchors, potentially causing over-merging and degrading accuracy.

\textbf{Robustness of Dynamic Settings.} In contrast, the dynamic setting (left part of Figure~\ref{fig:anchor-number}) demonstrates superior stability and generalization.
By scaling $k$ proportionally to the bucket size, \tool maintains a consistent balance between identifying dense groups and filtering sparse ones.
Specifically, increasing the ratio from $1/4$ to $1/2$ significantly reduces computational cost as more logs are successfully routed to the efficient statistical processor.
However, further increasing the ratio beyond $1/2$ yields diminishing returns in efficiency while slightly risking over-merging (evident in the slight drop in FTA at $3/4$).
Therefore, to strike the optimal balance between effectiveness and cost-effectiveness, we adopt the dynamic setting with a ratio of $1/2$ as our default configuration.

\begin{tcolorbox}
[
    colback=gray!5!white,
    colframe=gray!75!black,
    fonttitle=\bfseries,
    sharp corners
]
\vspace{-0.1in}
\small
\textbf{Answer to RQ3:}
\tool demonstrates minimal sensitivity to parameter settings. 
Both the dynamic similarity threshold and dynamic top-$k$ selection mechanisms ensure stable performance, with overall accuracy fluctuations of less than 1.5\% across a wide range of parameter settings.
\vspace{-0.1in}
\end{tcolorbox}

\begin{table}[tbp]
\centering
\footnotesize
\caption{\tool's performance with different LLMs.}
\label{tab:different_llms}
\begin{tabular}{lcccc}
\toprule
Model  & GA & PA & FGA & FTA \\
\midrule

\rowcolor{gray!20}\multicolumn{5}{c}{\textbf{Default}} \\ 
\midrule

GPT-5.2 & 0.927 & 0.875 & 0.873 & 0.773 \\

\midrule

\rowcolor{gray!20}\multicolumn{5}{c}{\textbf{Large Language Models}} \\
\midrule


Gemini-3-pro &  0.928 & 0.876 & 0.877  & 0.772  \\

\midrule

Qwen3-Max  & 0.924  & 0.868 & 0.870 & 0.764 \\

\midrule

Claude-sonnet-4.5  & 0.924  & 0.872  & 0.871  & 0.767 \\

\midrule
\rowcolor{gray!20}\multicolumn{5}{c}{\textbf{Small Language Models}} \\
\midrule

GPT-5-mini &  0.926  & 0.870 & 0.874  & 0.768 \\

\midrule

Gemini-2.5-flash-lite & 0.924  & 0.870  & 0.868 & 0.762 \\

\midrule


{Qwen3-32b}  & 0.924 & 0.862 & 0.865  & 0.756 \\


{Qwen3-14b}  & 0.915  & 0.861  & 0.870  & 0.749 \\

{Qwen3-8b}  & 0.915  & 0.863  & 0.863  & 0.744\\



\bottomrule
\end{tabular}
\end{table}
\subsection{RQ4: Robustness Analysis}
To evaluate the generalizability and robustness of \tool, we conduct extensive experiments across a diverse spectrum of backbone models, ranging from state-of-the-art Large Language Models like Gemini-3-pro to more efficient Small Language Models such as Qwen3-8b. 
Table~\ref{tab:different_llms} shows the results.
The results demonstrate that \tool maintains remarkably stable performance regardless of the underlying model's scale. For instance, the variance in GA remains minimal, with Gemini-3-pro achieving $0.928$ and the smaller Qwen3-8b still maintaining a high $0.915$.
This stability stems from our core hybrid architecture, which integrates statistical heuristics with LLM-based reasoning. 
By offloading deterministic patterns to the statistical component, we reduce the burden on the LLM, ensuring that even smaller models can deliver results comparable to their larger counterparts. 
Furthermore, the results also confirm the pivotal role of the proposed router in this robustness. 
It effectively steers log entries to the most appropriate processing path, preventing the system from over-relying on the reasoning capabilities of any single backbone. Even when utilizing Gemini-2.5-flash-lite, \tool achieves an FTA of $0.762$, which is nearly identical to the $0.772$ produced by the much larger Pro version. These findings confirm that \tool is not vertically dependent on a specific proprietary model but is a framework capable of delivering consistent, high-quality parsing results across various models.

\begin{tcolorbox}
[
    colback=gray!5!white,
    colframe=gray!75!black,
    fonttitle=\bfseries,
    sharp corners
]
\vspace{-0.1in}
\small
\textbf{Answer to RQ4:}
\tool demonstrates robustness across diverse backbone LLMs, maintaining stable performance from large language models to smaller scales.
\vspace{-0.1in}
\end{tcolorbox}

\section{Threats to Validity}
\textbf{Internal Threats.} 
The primary internal threat arises from the inherent randomness of the LLM used in the slow track. To mitigate this stochastic behavior, we set the temperature parameter to zero. This ensures the model generates deterministic outputs for identical inputs. Additionally, we repeated all experiments three times to average out any remaining fluctuations. Another internal threat involves the implementation bias of baseline methods. We addressed this by adopting the official open-source implementations for all comparison partners. We also maintained consistent hyperparameters to ensure a fair evaluation environment.

\textbf{External Threats.} 
A major external threat is the potential data leakage during the LLM training phase. Since LLMs consume vast amounts of public data, they might memorize log templates. However, our approach tasks the model with variable identification rather than full template generation. This specific instruction reduces the reliance on memorized patterns. Furthermore, the quality of the evaluation datasets impacts generalizability. To counter this, we conducted evaluations on the extensive LogHub-2.0 benchmark. These datasets span various system types and verify that our method generalizes well across different domains.
\section{Related Work}
\textbf{Log Parsing.} 
Log parsing aims to transform raw log messages into structured templates by identifying dynamic parameters~\cite{le2023log, jiang2024lilac,huang2025no,sedki2022effective,vaarandi2003data}. 
Existing methods are primarily categorized into syntax-based~\cite{he2017drain,dai2020logram,dai2023pilar,makanju2009clustering,fu2009LKE} and semantic-based approaches~\cite{huo2023semparser,liu2022uniparser,le2023log,ma2024librelog,zhang2025semanticlog}. 
The syntax-based parsers, such as Drain~\cite{he2017drain} and LogMine~\cite{hamooni2016logmine}, rely on heuristic rules or tree structures to extract frequent patterns. While efficient, they often lack flexibility when dealing with complex or evolving logs. 
Semantic-based methods usually utilize neural networks to capture log meanings. However, they depend heavily on labeled datasets, which are expensive to annotate. 
Recently, LLMs have achieved state-of-the-art performance in this task~\cite{jiang2024lilac,huang2025no,xiao2024free,ma2024librelog}. Despite their high accuracy, applying LLMs to every log message incurs prohibitive computational and financial costs~(i.e., longer parsing time, higher token consumption, and more invocations). To address this, our approach leverages the insight that not all logs require heavy LLM inference. We identify and categorize logs into dense and sparse groups. A dynamic router then directs dense groups to an efficient statistical processor and only forwards sparse groups to the LLM. This hybrid design effectively boosts high parsing accuracy with cost efficiency.

\textbf{Automated Log Analytics.}
Log parsing serves as a critical step for downstream tasks~\cite{chen2021experience,ali2023empirical,lou2010mining}, such as anomaly detection~\cite{du2017deeplog,zhang2019robust,he2025weakly,meng2019loganomaly}, root cause analysis~\cite{wang2020root,wittkopp2024logrca,cui2025aetherlog}, and system error localization~\cite{shan2024face,wang2018misconfdoctor}.
In the era of LLMs, many LLM-based tools targeted at automated log analytics have been proposed~\cite{shan2024face,notaro2023logrule,xu2025logsage,jiang2025logpilot,xiao2025clslog}.
These powerful tools excel at understanding complex system behaviors and diagnosing unseen errors. However, their practical deployment still relies on high-quality and structured log inputs. Our cost-efficient parser ensures that these downstream tasks receive accurate templates without excessive computational overhead.

\section{Conclusion}
In this paper, we propose \tool, an efficient and cost-effective log parser. Our work is driven by the insight that not all logs require complex semantic understanding. We design a dynamic routing mechanism to classify logs into dense and sparse groups. Since dense groups have clear statistical signals, we direct them to a fast statistical processor. This ensures stability and eliminates hallucination risks without incurring LLM costs. Conversely, sparse groups lack these patterns. Therefore, we route them to an LLM processor for semantic analysis. 
Extensive experiments on public datasets show the superiority of \tool. It achieves superior grouping and parsing accuracy compared to current baselines. Furthermore, \tool parses logs much faster than existing LLM methods and syntax parsers. Finally, it reduces token consumption by orders of magnitude, making it highly practical for real-world applications.

\bibliographystyle{ACM-Reference-Format}
\bibliography{main}

@article{huang2025no,
  title={No more labelled examples? an unsupervised log parser with llms},
  author={Huang, Junjie and Jiang, Zhihan and Chen, Zhuangbin and Lyu, Michael},
  journal={Proceedings of the ACM on Software Engineering},
  volume={2},
  number={FSE},
  pages={2406--2429},
  year={2025},
  publisher={ACM New York, NY, USA}
}

@inproceedings{xiao2024free,
  title={free: Towards more practical log parsing with large language models},
  author={Xiao, Yi and Le, Van-Hoang and Zhang, Hongyu},
  booktitle={Proceedings of the 39th IEEE/ACM International Conference on Automated Software Engineering},
  pages={153--165},
  year={2024}
}

@article{jiang2024lilac,
  title={Lilac: Log parsing using llms with adaptive parsing cache},
  author={Jiang, Zhihan and Liu, Jinyang and Chen, Zhuangbin and Li, Yichen and Huang, Junjie and Huo, Yintong and He, Pinjia and Gu, Jiazhen and Lyu, Michael R},
  journal={Proceedings of the ACM on Software Engineering},
  volume={1},
  number={FSE},
  pages={137--160},
  year={2024},
  publisher={ACM New York, NY, USA}
}

@inproceedings{jiang2024large,
  title={A large-scale evaluation for log parsing techniques: How far are we?},
  author={Jiang, Zhihan and Liu, Jinyang and Huang, Junjie and Li, Yichen and Huo, Yintong and Gu, Jiazhen and Chen, Zhuangbin and Zhu, Jieming and Lyu, Michael R},
  booktitle={Proceedings of the 33rd ACM SIGSOFT International Symposium on Software Testing and Analysis},
  pages={223--234},
  year={2024}
}

@inproceedings{shan2024face,
  title={Face it yourselves: An llm-based two-stage strategy to localize configuration errors via logs},
  author={Shan, Shiwen and Huo, Yintong and Su, Yuxin and Li, Yichen and Li, Dan and Zheng, Zibin},
  booktitle={Proceedings of the 33rd ACM SIGSOFT international symposium on software testing and analysis},
  pages={13--25},
  year={2024}
}

@inproceedings{zhu2023loghub,
  title={Loghub: A large collection of system log datasets for ai-driven log analytics},
  author={Zhu, Jieming and He, Shilin and He, Pinjia and Liu, Jinyang and Lyu, Michael R},
  booktitle={2023 IEEE 34th International Symposium on Software Reliability Engineering (ISSRE)},
  pages={355--366},
  year={2023},
  organization={IEEE}
}

@article{yu2023brain,
  title={Brain: Log parsing with bidirectional parallel tree},
  author={Yu, Siyu and He, Pinjia and Chen, Ningjiang and Wu, Yifan},
  journal={IEEE Transactions on Services Computing},
  volume={16},
  number={5},
  pages={3224--3237},
  year={2023},
  publisher={IEEE}
}

@inproceedings{le2021log,
  title={Log-based anomaly detection without log parsing},
  author={Le, Van-Hoang and Zhang, Hongyu},
  booktitle={2021 36th IEEE/ACM International Conference on Automated Software Engineering (ASE)},
  pages={492--504},
  year={2021},
  organization={IEEE}
}

@inproceedings{huo2023semparser,
  title={Semparser: A semantic parser for log analytics},
  author={Huo, Yintong and Su, Yuxin and Lee, Cheryl and Lyu, Michael R},
  booktitle={2023 IEEE/ACM 45th International Conference on Software Engineering (ICSE)},
  pages={881--893},
  year={2023},
  organization={IEEE}
}

@inproceedings{du2017deeplog,
  title={Deeplog: Anomaly detection and diagnosis from system logs through deep learning},
  author={Du, Min and Li, Feifei and Zheng, Guineng and Srikumar, Vivek},
  booktitle={Proceedings of the 2017 ACM SIGSAC conference on computer and communications security},
  pages={1285--1298},
  year={2017}
}

@inproceedings{zhang2019robust,
  title={Robust log-based anomaly detection on unstable log data},
  author={Zhang, Xu and Xu, Yong and Lin, Qingwei and Qiao, Bo and Zhang, Hongyu and Dang, Yingnong and Xie, Chunyu and Yang, Xinsheng and Cheng, Qian and Li, Ze and others},
  booktitle={Proceedings of the 2019 27th ACM joint meeting on European software engineering conference and symposium on the foundations of software engineering},
  pages={807--817},
  year={2019}
}

@String{Computing = "Computing" }

@String{Computer = "{IEEE} Computer" }

@String{Springer = "Springer-Verlag" }

@inproceedings{zhu2019tools,
  title={Tools and benchmarks for automated log parsing},
  author={Zhu, Jieming and He, Shilin and Liu, Jinyang and He, Pinjia and Xie, Qi and Zheng, Zibin and Lyu, Michael R},
  booktitle={2019 IEEE/ACM 41st International Conference on Software Engineering: Software Engineering in Practice (ICSE-SEIP)},
  pages={121--130},
  year={2019},
  organization={IEEE}
}

@inproceedings{vaarandi2003data,
  title={A data clustering algorithm for mining patterns from event logs},
  author={Vaarandi, Risto},
  booktitle={Proceedings of the 3rd IEEE Workshop on IP Operations \& Management (IPOM)(IEEE Cat. No. 03EX764)},
  pages={119--126},
  year={2003},
  organization={Ieee}
}

@inproceedings{liu2022uniparser,
  title={Uniparser: A unified log parser for heterogeneous log data},
  author={Liu, Yudong and Zhang, Xu and He, Shilin and Zhang, Hongyu and Li, Liqun and Kang, Yu and Xu, Yong and Ma, Minghua and Lin, Qingwei and Dang, Yingnong and others},
  booktitle={Proceedings of the ACM Web Conference 2022 (WWW)},
  pages={1893--1901},
  year={2022}
}

@inproceedings{vaarandi2015logcluster,
  title={Logcluster-a data clustering and pattern mining algorithm for event logs},
  author={Vaarandi, Risto and Pihelgas, Mauno},
  booktitle={2015 11th International conference on network and service management (CNSM)},
  pages={1--7},
  year={2015},
  organization={IEEE}
}

@article{dai2020logram,
  title={Logram: Efficient Log Parsing Using $ n $ n-Gram Dictionaries},
  author={Dai, Hetong and Li, Heng and Chen, Che-Shao and Shang, Weiyi and Chen, Tse-Hsun},
  journal={IEEE Transactions on Software Engineering (TSE)},
  volume={48},
  number={3},
  pages={879--892},
  year={2020},
  publisher={IEEE}
}

@inproceedings{fu2009LKE,
  title={Execution anomaly detection in distributed systems through unstructured log analysis},
  author={Fu, Qiang and Lou, Jian-Guang and Wang, Yi and Li, Jiang},
  booktitle={2009 ninth IEEE international conference on data mining (ICDM)},
  pages={149--158},
  year={2009},
  organization={IEEE}
}

@inproceedings{hamooni2016logmine,
  title={Logmine: Fast pattern recognition for log analytics},
  author={Hamooni, Hossein and Debnath, Biplob and Xu, Jianwu and Zhang, Hui and Jiang, Guofei and Mueen, Abdullah},
  booktitle={Proceedings of the 25th ACM International on Conference on Information and Knowledge Management (CIKM)},
  pages={1573--1582},
  year={2016}
}

@article{shima2016LenMa,
  title={Length matters: Clustering system log messages using length of words},
  author={Shima, Keiichi},
  journal={arXiv preprint arXiv:1611.03213},
  year={2016}
}

@inproceedings{wittkopp2024logrca,
  title={Logrca: Log-based root cause analysis for distributed services},
  author={Wittkopp, Thorsten and Wiesner, Philipp and Kao, Odej},
  booktitle={European Conference on Parallel Processing},
  pages={362--376},
  year={2024},
  organization={Springer}
}

@inproceedings{jiang2008abstracting,
  title={Abstracting execution logs to execution events for enterprise applications (short paper)},
  author={Jiang, Zhen Ming and Hassan, Ahmed E and Flora, Parminder and Hamann, Gilbert},
  booktitle={2008 The Eighth International Conference on Quality Software},
  pages={181--186},
  year={2008},
  organization={IEEE}
}

@inproceedings{makanju2009clustering,
  title={Clustering event logs using iterative partitioning},
  author={Makanju, Adetokunbo AO and Zincir-Heywood, A Nur and Milios, Evangelos E},
  booktitle={Proceedings of the 15th ACM SIGKDD international conference on Knowledge discovery and data mining (KDD)},
  pages={1255--1264},
  year={2009}
}

@inproceedings{he2017drain,
  title={Drain: An online log parsing approach with fixed depth tree},
  author={He, Pinjia and Zhu, Jieming and Zheng, Zibin and Lyu, Michael R},
  booktitle={2017 IEEE international conference on web services (ICWS)},
  pages={33--40},
  year={2017},
  organization={IEEE}
}

@article{le2023log,
  title={Log Parsing with Prompt-based Few-shot Learning},
  author={Le, Van-Hoang and Zhang, Hongyu},
  journal={arXiv preprint arXiv:2302.07435},
  year={2023}
}

@inproceedings{he2016evaluation,
  title={An evaluation study on log parsing and its use in log mining},
  author={He, Pinjia and Zhu, Jieming and He, Shilin and Li, Jian and Lyu, Michael R},
  booktitle={2016 46th annual IEEE/IFIP international conference on dependable systems and networks (DSN)},
  pages={654--661},
  year={2016},
  organization={IEEE}
}

@article{notaro2023logrule,
  title={LogRule: Efficient Structured Log Mining for Root Cause Analysis},
  author={Notaro, Paolo and Haeri, Soroush and Cardoso, Jorge and Gerndt, Michael},
  journal={IEEE Transactions on Network and Service Management},
  year={2023},
  publisher={IEEE}
}

@inproceedings{schipper2019tracing,
  title={Tracing back log data to its log statement: from research to practice},
  author={Schipper, Daan and Aniche, Maur{\'\i}cio and van Deursen, Arie},
  booktitle={2019 IEEE/ACM 16th International Conference on Mining Software Repositories (MSR)},
  pages={545--549},
  year={2019},
  organization={IEEE}
}

@article{chen2021experience,
  title={Experience report: Deep learning-based system log analysis for anomaly detection},
  author={Chen, Zhuangbin and Liu, Jinyang and Gu, Wenwei and Su, Yuxin and Lyu, Michael R},
  journal={arXiv preprint arXiv:2107.05908},
  year={2021}
}

@misc{openai-api,
  title={OpenAI API},
  howpublished = {\url{https://openai.com/blog/openai-api}},
  year=2026,
  note = "{Online; Accessed: 2026-03-16}"
}

@inproceedings{dai2023pilar,
  title={PILAR: Studying and Mitigating the Influence of Configurations on Log Parsing},
  author={Dai, Hetong and Tang, Yiming and Li, Heng and Shang, Weiyi},
  booktitle={2023 IEEE/ACM 45th International Conference on Software Engineering (ICSE)},
  pages={818--829},
  year={2023},
  organization={IEEE}
}

@inproceedings{yuan2010sherlog,
  title={Sherlog: error diagnosis by connecting clues from run-time logs},
  author={Yuan, Ding and Mai, Haohui and Xiong, Weiwei and Tan, Lin and Zhou, Yuanyuan and Pasupathy, Shankar},
  booktitle={Proceedings of the fifteenth International Conference on Architectural support for programming languages and operating systems},
  pages={143--154},
  year={2010}
}

@inproceedings{wang2018misconfdoctor,
  title={MisconfDoctor: diagnosing misconfiguration via log-based configuration testing},
  author={Wang, Teng and Liu, Xiaodong and Li, Shanshan and Liao, Xiangke and Li, Wang and Liao, Qing},
  booktitle={2018 IEEE International Conference on Software Quality, Reliability and Security (QRS)},
  pages={1--12},
  year={2018},
  organization={IEEE}
}

@inproceedings{wang2020root,
  title={Root-cause metric location for microservice systems via log anomaly detection},
  author={Wang, Lingzhi and Zhao, Nengwen and Chen, Junjie and Li, Pinnong and Zhang, Wenchi and Sui, Kaixin},
  booktitle={2020 IEEE international conference on web services (ICWS)},
  pages={142--150},
  year={2020},
  organization={IEEE}
}

@inproceedings{sedki2022effective,
  title={An Effective Approach for Parsing Large Log Files},
  author={Sedki, Issam and Hamou-Lhadj, Abdelwahab and Ait-Mohamed, Otmane and Shehab, Mohammed A},
  booktitle={2022 IEEE International Conference on Software Maintenance and Evolution (ICSME)},
  pages={1--12},
  year={2022},
  organization={IEEE}
}

@inproceedings{zhou2019latent,
  title={Latent error prediction and fault localization for microservice applications by learning from system trace logs},
  author={Zhou, Xiang and Peng, Xin and Xie, Tao and Sun, Jun and Ji, Chao and Liu, Dewei and Xiang, Qilin and He, Chuan},
  booktitle={Proceedings of the 2019 27th ACM Joint Meeting on European Software Engineering Conference and Symposium on the Foundations of Software Engineering (FSE)},
  pages={683--694},
  year={2019}
}

@inproceedings{shin2021theoretical,
  title={A theoretical framework for understanding the relationship between log parsing and anomaly detection},
  author={Shin, Donghwan and Khan, Zanis Ali and Bianculli, Domenico and Briand, Lionel},
  booktitle={International Conference on Runtime Verification},
  pages={277--287},
  year={2021},
  organization={Springer}
}

@inproceedings{liu2019logzip,
  title={Logzip: Extracting hidden structures via iterative clustering for log compression},
  author={Liu, Jinyang and Zhu, Jieming and He, Shilin and He, Pinjia and Zheng, Zibin and Lyu, Michael R},
  booktitle={2019 34th IEEE/ACM International Conference on Automated Software Engineering (ASE)},
  pages={863--873},
  year={2019},
  organization={IEEE}
}

@inproceedings{li2024logshrink,
  title={Logshrink: Effective log compression by leveraging commonality and variability of log data},
  author={Li, Xiaoyun and Zhang, Hongyu and Le, Van-Hoang and Chen, Pengfei},
  booktitle={Proceedings of the 46th IEEE/ACM International Conference on Software Engineering},
  pages={1--12},
  year={2024}
}

@inproceedings{yu2024unlocking,
  title={Unlocking the power of numbers: Log compression via numeric token parsing},
  author={Yu, Siyu and Wu, Yifan and Li, Ying and He, Pinjia},
  booktitle={Proceedings of the 39th IEEE/ACM International Conference on Automated Software Engineering},
  pages={919--930},
  year={2024}
}

@inproceedings{chen2018automated,
  title={An automated approach to estimating code coverage measures via execution logs},
  author={Chen, Boyuan and Song, Jian and Xu, Peng and Hu, Xing and Jiang, Zhen Ming},
  booktitle={Proceedings of the 33rd ACM/IEEE International Conference on Automated Software Engineering},
  pages={305--316},
  year={2018}
}

@article{ali2023empirical,
  title={An Empirical Study on Log-based Anomaly Detection Using Machine Learning},
  author={Ali, Shan and Boufaied, Chaima and Bianculli, Domenico and Branco, Paula and Briand, Lionel and Aschbacher, Nathan},
  journal={arXiv preprint arXiv:2307.16714},
  year={2023}
}

@misc{ibm-drain,
    author    = {David Ohana},
year ={2020},
  title = {A blog about Drain usage in IBM Cloud},
  howpublished = {\url{https://developer.ibm.com/blogs/how-mining-log-templates-can-help-ai-ops-in-cloud-scale-data-centers/}},
  note = {Online; Accessed: 2026-03-16}
}

@misc{spacy,
year ={2016},
  title = {Industrial-Strength Natural Language Processing},
  howpublished = {\url{https://spacy.io/}},
  note = {Online; Accessed: 2026-03-16}
}

@misc{nltk,
year ={2001},
  title = {NLTK Project},
  howpublished = {\url{https://www.nltk.org/howto/wordnet.html}},
  note = {Oneline; Accessed: 2026-03-16}
}

@book{bird2009natural,
  title={Natural language processing with Python: analyzing text with the natural language toolkit},
  author={Bird, Steven and Klein, Ewan and Loper, Edward},
  year={2009},
  publisher={" O'Reilly Media, Inc."}
}

@inproceedings{niwattanakul2013using,
  title={Using of Jaccard coefficient for keywords similarity},
  author={Niwattanakul, Suphakit and Singthongchai, Jatsada and Naenudorn, Ekkachai and Wanapu, Supachanun},
  booktitle={Proceedings of the international multiconference of engineers and computer scientists},
  volume={1},
  number={6},
  pages={380--384},
  year={2013}
}

@misc{gpt, 
author = {OpenAI},
year={},
title = {GPT-5.2},
howpublished = {\url{https://platform.openai.com/docs/models/gpt-5.2}}, 
note = {Oneline; Accessed: 2026-03-16}
}

@inproceedings{he2025weakly,
  title={Weakly-supervised log-based anomaly detection with inexact labels via multi-instance learning},
  author={He, Minghua and Jia, Tong and Duan, Chiming and Cai, Huaqian and Li, Ying and Huang, Gang},
  booktitle={2025 IEEE/ACM 47th International Conference on Software Engineering (ICSE)},
  pages={2918--2930},
  year={2025},
  organization={IEEE}
}

@inproceedings{lou2010mining,
  title={Mining invariants from console logs for system problem detection},
  author={Lou, Jian-Guang and Fu, Qiang and Yang, Shenqi and Xu, Ye and Li, Jiang},
  booktitle={2010 USENIX annual technical conference (USENIX ATC 10)},
  year={2010}
}

@inproceedings{cui2025aetherlog,
  title={AetherLog: Log-based Root Cause Analysis by Integrating Large Language Models with Knowledge Graphs},
  author={Cui, Tianyu and Fu, Ruowei and Liu, Changchang and Ji, Yuhe and Gu, Wenwei and Zhang, Shenglin and Sun, Yongqian and Pei, Dan},
  booktitle={2025 IEEE 36th International Symposium on Software Reliability Engineering (ISSRE)},
  pages={49--60},
  year={2025},
  organization={IEEE}
}

@inproceedings{meng2019loganomaly,
  title={Loganomaly: Unsupervised detection of sequential and quantitative anomalies in unstructured logs.},
  author={Meng, Weibin and Liu, Ying and Zhu, Yichen and Zhang, Shenglin and Pei, Dan and Liu, Yuqing and Chen, Yihao and Zhang, Ruizhi and Tao, Shimin and Sun, Pei and others},
  booktitle={Ijcai},
  volume={19},
  number={7},
  pages={4739--4745},
  year={2019}
}

@article{ma2024librelog,
  title={Librelog: Accurate and efficient unsupervised log parsing using open-source large language models},
  author={Ma, Zeyang and Kim, Dong Jae and Chen, Tse-Hsun},
  journal={arXiv preprint arXiv:2408.01585},
  year={2024}
}

@article{xu2025logsage,
  title={LogSage: An LLM-based framework for CI/CD failure detection and remediation with industrial validation},
  author={Xu, Weiyuan and Luo, Juntao and Huang, Tao and Sui, Kaixin and Geng, Jie and Ma, Qijun and Akasaka, Isami and Shi, Xiaoxue and Tang, Jing and Cai, Peng},
  journal={arXiv preprint arXiv:2506.03691},
  year={2025}
}

@article{zhang2025semanticlog,
  title={SemanticLog: Towards Effective and Efficient Large-Scale Semantic Log Parsing},
  author={Zhang, Chenbo and Xu, Wenying and Liu, Jinbu and Zhang, Lu and Liu, Guiyang and Guan, Jihong and Zhou, Qi and Zhou, Shuigeng},
  journal={IEEE Transactions on Software Engineering},
  year={2025},
  publisher={IEEE}
}

@article{jiang2025logpilot,
  title={LogPilot: Intent-aware and Scalable Alert Diagnosis for Large-scale Online Service Systems},
  author={Jiang, Zhihan and Liu, Jinyang and Li, Yichen and Huang, Haiyu and He, Xiao and Zhang, Tieying and Chen, Jianjun and Li, Yi and Shi, Rui and Lyu, Michael R},
  journal={arXiv preprint arXiv:2509.25874},
  year={2025}
}

@inproceedings{xiao2025clslog,
  title={Clslog: Collaborating large and small models for log-based anomaly detection},
  author={Xiao, Pei and Jia, Tong and Duan, Chiming and He, Minghua and Hong, Weijie and Yang, Xixuan and Wu, Yihan and Li, Ying and Huang, Gang},
  booktitle={Proceedings of the 33rd ACM International Conference on the Foundations of Software Engineering},
  pages={686--690},
  year={2025}
}

\end{document}